\documentclass[11pt]{article}
\usepackage[utf8]{inputenc}
\usepackage{graphicx}
\usepackage[frozencache,cachedir=.]{minted}
\usepackage[linesnumbered,ruled,noend,vlined]{algorithm2e}
\usepackage{hyperref}    

\usepackage{boxedminipage}
\newenvironment{result}
{\smallskip
\noindent
\let\emph=\textbf
\begin{boxedminipage}{\columnwidth}\em}
{\end{boxedminipage}
}

\newtheorem{definition}{Definition}

\newcommand{\rest}[1]{REST API#1}
\newcommand{\rests}[1]{REST APIs#1}
\newcommand{\restful}[1]{RESTful API#1}
\newcommand{\restfuls}[1]{RESTful APIs#1}

\usepackage{geometry}
\title{Exploring Behaviours of RESTful APIs in an Industrial Setting}

\author{Stefan Karlsson\textsuperscript{1,2}, Robbert Jongeling\textsuperscript{2}, Adnan \v{C}au\v{s}evi\'{c}\textsuperscript{1}, Daniel Sundmark\textsuperscript{2}\\ \textsuperscript{1}ABB AB \\ \{stefan.l.karlsson, adnan.causevic\}@se.abb.com  \\ \textsuperscript{2} Mälardalen University \\ \{stefan.l.karlsson, daniel.sundmark, robbert.jongeling\}@mdu.se }
\date{}

\begin{document}

\maketitle

\begin{abstract}

A common way of exposing functionality in contemporary systems is by providing a Web-API based on the \rest{} architectural guidelines. To describe \rests{}, the industry standard is currently OpenAPI-specifications. Test generation and fuzzing methods targeting OpenAPI-described \rests{} have been a very active research area in recent years. An open research challenge is to aid users in better understanding their API, in addition to finding faults and to cover all the code. In this paper, we address this challenge by proposing a set of behavioural properties, common to \rests{}, which are used to generate examples of behaviours that these APIs exhibit. These examples can be used both (i) to further the understanding of the API and (ii) as a source of automatic test cases. Our evaluation shows that our approach can generate examples deemed \emph{relevant} for understanding the system and for a source of test generation by practitioners. In addition, we show that basing test generation on behavioural properties provides tests that are less dependent on the state of the system, while at the same time yielding a similar code coverage as state-of-the-art methods in \rest{} fuzzing in a given time limit.
\end{abstract}


\section{Introduction}

Web technology is a common way for systems to expose software functionality to clients, often following the architectural style of REST~\cite{fielding-REST-2000}. Services exposing REST Application Programming Interfaces (APIs) are common both for publicly available services, such as those provided by Google~\cite{Google-cloud}, Amazon~\cite{AWS-cloud} and Microsoft~\cite{Azure-cloud}, and for internal systems, in a microservice-based architecture~\cite{Fowler-Microservices}. To enable developers to use an API, some kind of description must be provided. Such descriptions can be provided in natural language, as written documentation, or in a more formally specified format.
The \emph{de facto} method of providing a specification---if any is provided---is by using OpenAPI-specifications~\cite{openapi-url}.
An OpenAPI-specification describes the exposed operations, their parameters and responses. 
The adoption of OpenAPI specifications is widespread and increasing~\cite{Serbout-Web-APIs-Structures-and-Data-Models-Analysis-2022}.

Given the popularity of \rests{}, methods to assess the quality of \rests{} have been of large interest in the research literature, with many different testing methods proposed~\cite{Arcuri-RESTful-API-Automated-Test-Case-Generation-with-EvoMaster-2019, Ed-douibi-Automatic-Generation-of-Test-Cases-for-REST-APIs-A-Specification-Based-Approach-2018, Atlidakis-RESTler-Stateful-REST-API-Fuzzing-2019, Karlsson-QuickREST-Property-based-Test-Generation-of-OpenAPI-Described-RESTful-APIs-2020, Viglianisi-RESTTESTGEN-Automated-Black-Box-Testing-of-RESTful-APIs-2020, Martin-Lopez-RESTest-Automated-Black-Box-Testing-of-RESTful-Web-APIs-2021, Laranjeiro-A-Black-Box-Tool-for-Robustness-Testing-of-REST-Services-2021, Wu-Combinatorial-Testing-of-RESTful-APIs-2022, Atlidakis-Pythia-Grammar-Based-Fuzzing-of-REST-APIs-2020}. The proposed methods cover a range of approaches, for example, search-based, property-based, and model-based techniques~\cite{Kim-Automated-Test-Generation-for-REST-APIs-No-Time-to-Rest-Yet-2022}. A common factor of the proposed methods is the reliance on an OpenAPI-specification, describing the operations of the system-under-test (SUT). Thus, OpenAPI-specifications are central to the current state-of-the-art (SotA) in test generation and fuzzing of \rests{}.

The common evaluation metrics in REST API test generation are fault finding and code coverage. According to recent studies~\cite{Zhang-Open-Problems-in-Fuzzing-REST-2022, Kim-Automated-Test-Generation-for-REST-APIs-No-Time-to-Rest-Yet-2022}, the currently best performing method---both in terms of code coverage~\cite{Zhang-Open-Problems-in-Fuzzing-REST-2022, Kim-Automated-Test-Generation-for-REST-APIs-No-Time-to-Rest-Yet-2022} and fault finding~\cite{Kim-Automated-Test-Generation-for-REST-APIs-No-Time-to-Rest-Yet-2022}---is EvoMaster~\cite{Arcuri-RESTful-API-Automated-Test-Case-Generation-with-EvoMaster-2019}. 
Fuzzing with EvoMaster produces on average about 50\% code coverage, on a set of different SUTs~\cite{Zhang-Open-Problems-in-Fuzzing-REST-2022, Kim-Automated-Test-Generation-for-REST-APIs-No-Time-to-Rest-Yet-2022}. However, the variability in coverage for different SUTs is quite high, ranging from about 15\% in the worst case and 95\% in the best case. These numbers are encouraging, for such a recent research area as REST API test generation. Regarding fault finding, a common measure of fault finding is to judge the API returning a 500 HTTP status-code (meaning ``Internal Server Error'') as a fault~\cite{Golmohammadi-Testing-RESTful-APIs-A-Survey-2023}.
However, aside from reaching high code coverage and fault finding, understanding how the API behaves is an important open challenge and essential to delivering required functionality of an API.

OpenAPI specifications include the operations, parameters, and responses of APIs. These specifications have enabled the automated generation of test cases that check the responses of the system with respect to the specified response or crashes. However, these tests do not provide sufficient information to users who want to use and understand the API. To gain this understanding, usage examples are an important resource~\cite{Robillard-What-Makes-APIs-Hard-to-Learn?-Answers-from-Developers-2009, Robillard-A-field-study-of-API-learning-obstacles-2011, McLellan-Building-more-usable-APIs-1998, Nykaza-What-Programmers-Really-Want:-Results-of-a-Needs-Assessment-for-SDK-Documentation-2002, Shull-Investigating-reading-techniques-for-object-oriented-framework-learning-2000, Novick-What-Users-Say-They-Want-in-Documentation-2006}.
In addition, relationships can exist between specified API operations. 
For example, the operations might have to be invoked in a specific order, or that one operation affects the output of another operation. If relationships between distinct API operations are unclear, discovering those can put a large burden on the user~\cite{Piccioni-An-Empirical-Study-of-API-Usability-2013}. Examples are of great value for understanding the behaviour of an API, thus, automatically generating such examples is valuable. In addition, if the generation can leverage already available artefacts---an OpenAPI specification for REST APIs---the required user effort is low.

In this paper, we primarily target the challenge of generating relevant examples of RESTful API behaviours to aid users in better understanding the behaviours of the API and to leverage such examples for test generation.
Using generated relevant examples for test generation addresses several aspects in REST API test generation, including (i) generating tests closer related to business logic, (ii) reducing the time of test generation, and (iii) decreasing the dependency on a clean state of the SUT in test generation and execution. All three of these have been identified as open research challenges~\cite{Zhang-Fuzzing-Microservices-In-Industry-2022, Kim-Automated-Test-Generation-for-REST-APIs-No-Time-to-Rest-Yet-2022}.
The approach we propose uses behavioural properties, common to \rests, to generate examples of behaviours. These examples are used as a source for test cases, and also as a source to further practitioners understanding of the API. Our approach, focusing on examples of behaviours complements approaches solely focusing on code coverage and fault finding---a very important purpose---which generate tests only furthering those goals. Consider an example of input validation, where an API operation \textit{does not} put any constraints on the input, i.e., there is no input validation code to cover. To fulfil the criteria of code coverage, we only have to execute the operation once. However, such a test may not show the actual behaviour when looking at the produced test case; the \textit{absence} of input validation. Focusing on behaviours opens up the possibility of generating tests for such scenarios, moving test generation closer to the actual business logic of \rests{}.

To evaluate our approach of using generated examples based on behavioural properties for \rest{} test-generation, we have extended the Property-based \rest{} test-generation method QuickREST~\cite{Karlsson-QuickREST-Property-based-Test-Generation-of-OpenAPI-Described-RESTful-APIs-2020}. Our main claim with the proposed approach is that by using an example generation approach, users can \emph{both} generate a source of understanding the SUT and a source for test generation---with comparable coverage to the SotA. In order to support this claim, we provide the results of two evaluations. First, we evaluate if our proposed approach provide practitioners with relevant examples of behaviours which are useful and aid in understanding the behaviour of the SUT. We do so by performing focus group sessions with experienced industry practitioners. Secondly, we use QuickREST with our extension and compare to EvoMaster, which is the current SotA for test generation driven by code coverage and fault finding. We evaluate the search coverage of our behaviour-based method, and also how the execution coverage of the resulting artefact behaves, in relation to the search coverage. Using code coverage as an evaluation metric is useful since a strong correlation between coverage and fault finding has been shown when evaluating different methods of REST API test generation~\cite{Kim-Automated-Test-Generation-for-REST-APIs-No-Time-to-Rest-Yet-2022}.

Our approach produces a set of examples of behaviours that can be used both as a guide to understand the system, and as a test suite. Our evaluation shows that practitioners deem these generated examples as relevant and see multiple usage areas for them. At the same time, using the generated examples as automatic test cases yields a code coverage comparable to the SotA.
In addition, test suites resulting from our approach are more robust to changes in the SUT state, making them more useful for regression testing, compared to test cases based on more specific expectations than general behaviours.
In summary, the contribution we make with this paper is a novel approach of \rest{} test generation, based on a set of proposed common behaviours of \rests{} producing relevant examples. This approach produces tests intending to cover behaviours, closer to the business logic of a system and provide API usage examples, which in addition can also be used to further the understanding of the SUT.

The structure of this paper is as follows; in Section~\ref{sec:key-ideas} we outline the key idea of the proposed approach. In Section \ref{sec:approach} we introduce our approach in detail. This is followed by the evaluation of the approach in two parts. In Sections~\ref{sec:evaluation-relevance}, we present the evaluation of the approaches ability to generate relevant examples. The evaluation of test-generation is in Section~\ref{sec:evaluation-test-gen}. In Section~\ref{sec:discussion} we discuss the results. The paper is then concluded with related work, Section \ref{sec:related-work}, and finally conclusions, Section \ref{sec:conclusions}.
\begin{figure*}
    \centering
    \includegraphics[width=\textwidth]{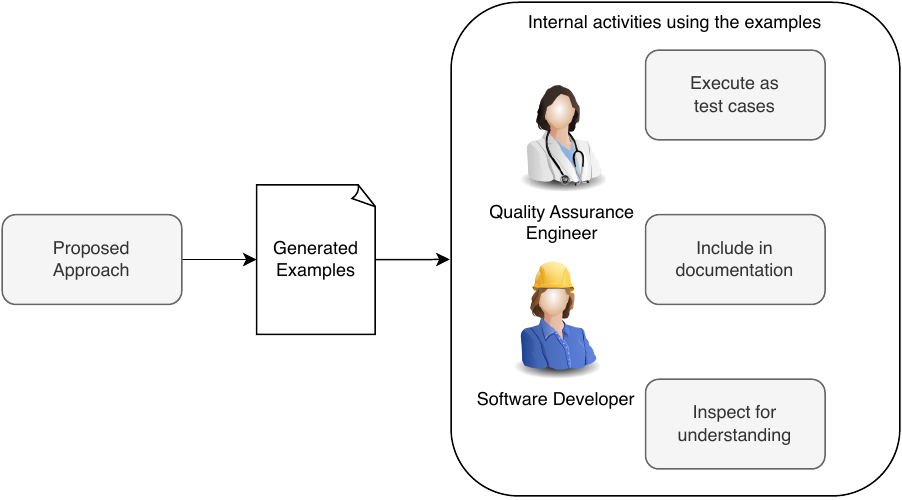}
    \caption{Overview of how the generated examples can be used to support internal development activities}
    \label{fig:concept-overview}
\end{figure*}

\section{Key Idea: Generating Examples of REST API Behaviours}\label{sec:key-ideas}

The key idea of our proposed approach is to use example generation as a source of understanding and test generation. The approach we propose generates examples of common \rest{} behaviours, such as managing the entities of the exposed system by creating, reading, and deleting them. We define these common behaviours as properties, for example, stating that a sequence of operations should incur a state-change in the system. By ``examples'', we refer to sequences of API calls that show a particular behaviour. In this paper, we generate such examples as sequences of REST API calls that show particular common REST API behaviours. Since the examples are sequences of API calls, they can be leveraged as tests themselves, or simply as a source of understanding the API, or as documentation for engineers/users of the API. Figure~\ref{fig:concept-overview} shows an overview of possible ways of working with generated examples.

To aid in the understanding of an API, generated examples should only include those API operations needed to show the exemplified behaviours, in other words, the examples must be \emph{relevant}. Relevant examples should include only those API operations that are important to the behaviour that the interaction exemplifies~\cite{Gerdes-Understanding-Formal-Specifications-through-Good-Examples-2018, Robillard-A-field-study-of-API-learning-obstacles-2011, Karlsson-Exploring-API-Behaviours-Through-Generated-Examples-2023}. 
In addition, Gerdes et al.~\cite{Gerdes-Understanding-Formal-Specifications-through-Good-Examples-2018} found that code coverage is not a strong heuristic for the relevance of generated examples. 
Thus, a key idea of our approach is to focus generation on finding behaviours, rather than metrics such as fault finding and code coverage.

Our proposed approach builds upon a general API example generation approach, introduced by Karlsson et al.~\cite{Karlsson-Exploring-API-Behaviours-Through-Generated-Examples-2023}. We adapt the general approach and put it in the context of REST API test generation. By doing so, we can generate examples of common REST API behaviours which can be used as test cases. In particular, we generate specific examples for REST APIs that can be both used as a source of information on the API's behaviour and as a source for test generation. 
The approach can be used as a complement to existing fuzzing methods, producing a broader range of test cases closer to the business logic of the application.

To generate examples of \rests{} behaviour, we define and use \textit{behavioural properties} common to RESTful APIs. These behaviours are based on the CRUD (\emph{Create, Read, Update, Delete}) idiom, typical for APIs managing entities. We use an OpenAPI specification to know which operations the API provides. We then leverage test generation to generate trial sequences, execute those on the system under test, and judge if the sequence matches any of the defined behavioural properties. This is done in a black-box fashion, since the only required input to the approach is the OpenAPI specification.

Specifically targeting behaviours of RESTful APIs provides an opportunity to leverage information specific to this API domain in the example generation process. Such information is the common CRUD idioms of RESTful APIs, and the expectations of the purpose of the HTTP methods, such as GET, POST, and DELETE. 

In using our example generating approach, we can produce test cases that cover behaviour, but that do not necessarily contribute to other metrics, such as code coverage.

\begin{figure*}
    \centering
    \includegraphics[width=\textwidth]{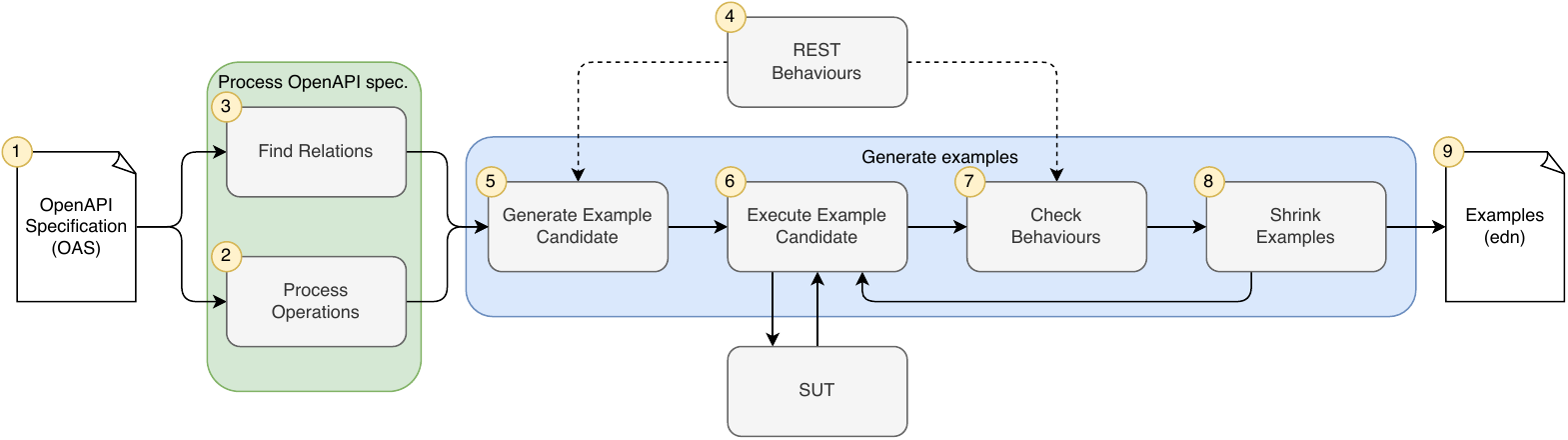}
    \caption{Overview of the example generation process}
    \label{fig:overview}
\end{figure*}

\section{Proposed Approach}\label{sec:approach}

In this section, we describe the details of our proposed approach. We explain how \textit{behavioural properties} are used to generate examples as test-cases for \rests{}. In doing so, we present the overarching algorithm and the specific properties formulated. Further, we present our solution for how to relate different operations in an OpenAPI specification (OAS). 

Figure~\ref{fig:overview} shows an overview of the approach. On a high-level, the input to the process is an OAS\includegraphics[scale=0.6]{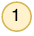}. We first process the OAS to know which operations are available\includegraphics[scale=0.6]{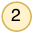} and how they relate\includegraphics[scale=0.6]{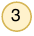}. This information is then used in conjunction with defined REST behaviours\includegraphics[scale=0.6]{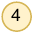} to generate examples. Generated example sequences\includegraphics[scale=0.6]{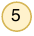} are executed on the SUT\includegraphics[scale=0.6]{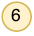} to observe how the SUT behaves in context of the example sequence\includegraphics[scale=0.6]{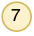}. The output is the shrunk\includegraphics[scale=0.6]{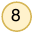} sequences of operations that passed a check of a defined behaviour\includegraphics[scale=0.6]{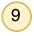}. The following sections will describe these parts in more detail.

A user of this approach can produce test cases that exemplify behaviours of the SUT that the user might expect, or be surprised by. In some cases the SUT might not expose some of the behaviours the approach can find.  In those cases, a lack of examples is the expected output---when the SUT does not express the given behaviour, no such examples should be produced. This method can not only be used for quality assurance purposes. The method is also applicable in cases where a user wants to learn more about an API---where examples may be beneficial~\cite{Robillard-A-field-study-of-API-learning-obstacles-2011}. Perhaps a developer is using a 3rd party API which lacks documentation. By generating behavioural-based test cases, the user can find out how this particular system behaves with respect to the behavioural properties defined.

\subsection{Defining Behaviours}

We use the term \textit{behaviour} in this paper. We use the term to focus on what an API should or should not do and to distinguish from \textit{faults}. A correct program will be free from faults, but it will have behaviours---behaviours are the reason a program exists. We use the term behaviour as describing a sequence of operations, with parameters if needed, that when executed on the SUT produces an observable nominal effect. The observable effect can be of different kinds. For example, one effect of executing a sequence of operations could be that we get an equal response from each operation, or the responses are different, or the sequence affects the response of another operation, etc. In a black-box scenario, we can only observe what the SUT returns. In the case of \rests{}, this is the status code and potentially data contained in the response.

A definition of a behavioural property serves as a general model in which a \rest{} specific behaviour is expressed. We summarize the description of a \textit{Behavioural Property} in Definition \ref{def:property}.
\begin{definition}[Behavioural Property]\label{def:property}
A \textit{Behavioural Property} is a property, $B=\langle C(O),Q \rangle$, defining a tuple containing a predicate $C$, based on the observations, $O$, determining if the effects of executing a sequence of operations, generated from $Q$, conforms to a defined pattern.
\end{definition}

Given our definition of behaviour in mind, the question is then; what sequences of operations should be generated and what specific effect-patterns to check?

\subsection{Definition of REST API-specific Behavioural Properties}

The knowledge that can be derived from a \rest{} is, for example, the meaning of the HTTP-methods used to invoke the API. An example of such knowledge would be an operation declared as an HTTP DELETE method---it can be derived that the intent of the operation is to delete an entity. Thus, when stating behaviours we can leverage that, for example, a GET-operation is expected to query entities and that DELETE-operations are expected to delete entities. This knowledge can be used both in the generation of sequences of operations to execute and in the conformance predicate checks of the observed effects of the execution of a sequence. In sequence generation, we can use this knowledge to set different probabilities of types of operations in the generated sequence---for example, for some behaviours it is more likely that a delete operation follows a create operation---or to select specific types of operations in certain positions in the sequence. For example, if we want to generate tests for a behaviour of successful deletion, it is pointless to generate a DELETE-operation prior to a POST-operation (create) of an entity. In the conformance checks of the observed effects, we can leverage REST-specific knowledge, such as a GET-operation (query) should not change the state of the SUT, while a successful DELETE-operation should. 

We define behavioural properties for \rests{} as: 
\begin{definition}[\restful{} Behavioural Property]\label{def:rest-property}
A \restful{} Behavioural Property, $B=\langle C(O), Q \rangle$, is a tuple including a predicate, $C$, which judge if a sequence of observations, $O$ conforms to an expected \restful{} behaviour, a generator $Q$ which generates execution sequences of HTTP-methods of a \rest{}, according to some selection rule.
\end{definition}

We consider an example of using this definition in which the generator $Q$ is defined to generate sequences of (i) first a GET-operation, (ii) any number of POST-operations, and (iii) the same GET-operation as started the sequence. When a sequence of API-calls generated by $Q$ is executed, the results returned by the API, the operation response and the status-code, are contained in $O$. After executing the API-call sequence, $O$ is judged by the boolean function $C$. In this example, $C$ will check if the observation $O_1$ is not equal to observation $O_n$, i.e., the observation of the first GET in the sequence and the last GET in the sequence should be different. If $C$ yields a positive result the observations is judged to conform to the behaviour, otherwise not.

There is, probably, an infinite set of possible behaviours of an API. Therefore, in practice, we must restrict ourselves to a finite set. A common idiom for \rests{} is to provide operations to manipulate entities, \textit{Create-, Read-, Update-, and Delete}-operations (CRUD). Thus, targeting CRUD would be a reasonable set of \rest{} behaviours to specify, which is what we do in this work. The specifics of these selected behaviours are expanded in Section \ref{sec:rest-behaviour-properties}.

\subsection{Exploration with Property-based Testing}

In this paper, we propose a method to explore the behavioural properties of a \rest{} by generating examples. As defined in Definition \ref{def:property}, we do this with the components of a sequence generator and a predicate based on the observations of executing the sequence. These components fit well into Property-based testing (PBT)~\cite{Claessen-QuickCheck-2000}. PBT libraries provide generators for basic types of values and some execution layer to run a specific number of trials when evaluating a property by random generation. However, these libraries typically only provide basic generators---such as for basic data types as strings, integers, etc.---and combinators, and the challenge of creating compositions of domain-specific generators and predicates are left to the user of the library.

PBT has been used in the context of \rests{}. Prior work by Karlsson et al. has shown that PBT is a method useful for automatic fault finding of OpenAPI-described \restfuls{}~\cite{Karlsson-QuickREST-Property-based-Test-Generation-of-OpenAPI-Described-RESTful-APIs-2020}. The focus of the work in \cite{Karlsson-QuickREST-Property-based-Test-Generation-of-OpenAPI-Described-RESTful-APIs-2020} was on fault finding, the proposed properties used were aimed at finding crashes (500 status codes in the case of \rests), and the conformance to the OpenAPI-specification of the actual responses of the SUT. In this work we take a different approach. While we still keep properties of fault finding---those are certainly of great importance---we also use behaviour-based properties, as described. We focus on, in addition to fault finding, generating tests to show behaviours (or the lack thereof) expected of a \restful{}. The main consequences of this is that the generation of operation sequences will depend on what behaviour we are trying to generate examples of. Moreover, the predicated checks do not only consider faults, but what defines the sought behaviour.

\subsection{Example-generation Algorithm}

\begin{algorithm}[t]\label{algo:exploration}
\SetKwProg{Let}{let}{:}{}
\SetKw{LetA}{\textbf{let}}
\SetKw{Ret}{\textbf{return}}
\SetKw{If}{\textbf{if}}
\SetKw{Then}{\textbf{then}}
\SetKw{Else}{\textbf{else}}
\SetKw{ElseIf}{\textbf{else if}}
\SetKw{Cont}{\textbf{continue}}
\caption{Example Generation}

\Let{Explore($B$, $S$, $N$)}{
  $os \gets QueryOperations(S)$\\
  $rs \gets BuildSchemaGraph(S)$\\
  $B_Q \gets BuildOperationsGenerator(B, os, rs)$\\
  $B_C \gets BuildBehviourCheck(B, os)$\\
  $Example \gets CheckProperty(B, N)$
}

$ $

\Let{CheckProperty($B$, $N$)} {
\For {$n \in 1..N$}{
  $(result, E) \gets CheckBehaviour(B_Q, B_C)$\\
  \If $result$
  \Then $Shrink(E)$\\
  \ElseIf $n = N$
    \Ret $\mathit{NoExampleFound}$\\
  \Else \Cont
}
}

$ $

\Let{CheckBehaviour($Q$, $C$)} {
$ E \gets Q_{GenerateOperationsSequence()}$\\
\For{$e \in E$}{
    $o \gets ExecuteOperation(e)$\\
    $\mathcal{O} \gets \mathcal{O} \cup ProcessObservation(o)$\\
}
$\mathcal{O}_{final} \gets ProcessObservations(\mathcal{O})$\\
\Ret $(C_{Check(E, \mathcal{O}, \mathcal{O}_{final})}, E)$\\

}

\end{algorithm}

In accordance with Definition~\ref{def:rest-property}, a behaviour consists of a check-function, $C$, over the observations, $O$, and an operation sequence generator function, $Q$. Thus, the first step of the exploration approach is to create those components based on the specific behaviour sought and the given API specification.

Referring to Algorithm~\ref{algo:exploration}, the two main inputs to our exploration approach are the OAS, $S$, providing a schema of the operations, parameters, and responses, and the specific behaviour, $B$, we are seeking. The number of iterations to search for an example before giving up is also an input, $N$.

We start by querying the operations of the OAS and build a graph of how the operations relate. The generator for the specific behaviour, $B_Q$, is then created based on the available operations in the OAS and their relations, in accordance with the sought behaviour. For example, if we seek a ``delete'' behaviour, we want the generator created to produce operation sequences with probable create operations of an entity (POST) before any deletion of the entity (DELETE). Which specific create and delete operations that can be generated as a trial example is based on how the operations, $os$, relate to each other, $rs$. The details of how relations are found are described further in section~\ref{sec:relation-finding}. The construction of the check function, $B_C$, also depends on the behaviour we seek. Using the same ``delete'' example as above, the check function created for such behaviour might verify that the observed behaviour first contains an entity, then later in the sequence of observed responses from the SUT, no longer included the same entity. Hence, the entity has been deleted and the sequence would be an example conforming to the sought behaviour. With the parts required for the exploration assembled, the exploration process is started, with a maximum of $N$ trials. This will produce an $\mathit{Example}$, or in the case of reaching the trial limit, a marker for $\mathit{NoExampleFound}$.

It is worth noting, given the approach of our method, a $\mathit{NoExampleFound}$ result for a given behaviour is not necessarily a bad thing. It can be the exact output we need to indicate that the SUT behave as we expect! For example, imagine a ``create'' operation allowing the creation of entities with the same parameters. If we seek examples of the behaviour where providing the same input twice results in a rejection of the second invocation, we will not find any examples---if the input validation is correct. Hence, not finding an example indicate expected behaviour, and found examples are potential bugs.

Considering the inner details of the algorithm (L8-21). We make $N$ trials to find an example of the behaviour. Each trial of checking for a behaviour starts by generating a sequence of operations to execute (L16), $E$, using the previously created generator and relations. Thus, each trial is a new sequence. Each operation in the sequence is executed on the SUT and the response of the execution is added to the set of observed responses. Before the observation is added, there is a possibility to process it. An example of such processing is that we might only be interested in parts of the response, such as the body and not headers containing meta-data. The final result of the check for behaviour is boolean, whether the sequence of execution in combination with the observations corresponds to the sought behaviour. If we do get a positive example of the behaviour (L11), we try to shrink the execution sequence to an example as short as possible that still exhibits the behaviour.

In summary, the input to the exploration method is the sought behaviour and the OAS, the output is either a shrunken example exhibiting the behaviour or $\mathit{NoExampleFound}$.

\subsection{Proposed REST API Behavioural Properties}\label{sec:rest-behaviour-properties}

When generating examples of general behaviours, Karlsson et al. defines and uses a set of \emph{meta-properties}~\cite{Karlsson-Exploring-API-Behaviours-Through-Generated-Examples-2023}. Meta-properties define general behaviours, such as detecting a state-change. The behaviours we propose in this paper are specific to REST APIs. In doing so, we instantiate general meta-properties and specialise them given our context.

In this first set of proposed REST API behavioural properties, we limit the set of behaviours to CRUD-based operations. In doing so, we take advantage of expectations around the HTTP verbs GET (``read''), POST (``create''), PUT/POST (``update''), and DELETE (``delete''). 
Recall from Definition~\ref{def:rest-property}, a Behaviour is a combination of a Check function and a generator of HTTP-methods. When we define a specific behaviour we then consider both the generation of reasonable trials and the check on them, these components are not separate. For example, if the behaviour we are seeking is deletion, we can, by construction, make sure the generator used will always include DELETE operations in the generated sequences. Any other sequences are not relevant to use as trials.

The set of initial behaviours we define and evaluate are:
\begin{itemize}
    \item \textbf{B1 - Equal response, same operation} - This behaviour generates a sequence of two invocations of the same operation with the same parameters. The range of examples this behaviour can produce is, for example, POST operations lacking input validation/restrictions, examples of GET operations reading the state of the SUT, or DELETE on an entity that does not exist, etc. Figure~\ref{fig:B1-example} shows two examples generated by this behaviour. The first example adds a product with the same name twice, i.e., enforcing unique names is not the current behaviour of this SUT. This example might confirm or surprise a developer/tester's perception of the SUT. The second example shows a GET operation returning the same response given the same parameters, and no other operation executed in between. Not finding such examples for GET operations would be a cause for further investigation.
    
    \item \textbf{B2 - Different response, same operation} - This behaviour generates a sequence of two invocations of the same operation, just as B1. However, the check of this behaviour expects the responses to be \emph{different}. The examples produced can be thought of as inverted from B1. For example, if the SUT constrained the ``productName'' from the B1 example to be unique, it would show as an example of this behaviour. The first response would have a successful status-code (such as 201) and the second response would be a rejection (400, indicating a client error). Any examples of GET operations generated by this behaviour would indicate a deviation from how a user would expect a typical CRUD API to behave.
\end{itemize}

\begin{figure}[h]
\begin{minted}[frame=single,linenos, xleftmargin=8pt, framesep=1mm, fontsize=\scriptsize, numbersep=2pt]{clojure}
{"addProduct"
  {:operation-sequence
   [{:operation "addProduct",
     :parameters 
     [{:name "productName", :value {"productName" "0"}}]}
    {:operation "addProduct",
     :parameters
     [{:name "productName", :value {"productName" "0"}}]}]
  }
}
     
{"getAllProducts"
  {:operation-sequence
   [{:operation "getAllProducts", :parameters []}
    {:operation "getAllProducts", :parameters []}]
  }
}
\end{minted}
\caption{Examples generated from behaviour B1}
\label{fig:B1-example}
\end{figure}

The defined behaviours B1 and B2 are simple; the sequence to test only consists of two operations. This is a trade-off between the complexity of behaviours possible to find and the duration of the search. A user making a change to input validation can choose to execute these behavioural properties, to get fast results on the behaviour currently under development. Still, while being simple, B1 and B2 have the ability to find several of the behaviours we expect of a CRUD API, as described, but are unable to find behaviours such as a successful deletion of a created entity. To get more complex examples, where the sequence of operations is not as constrained, we define two more behaviours. In the definitions, we categorize the HTTP-methods POST, DELETE, and PUT as potentially state-changing.
\begin{itemize}
    \item \textbf{B3 - A sequence of state-changing operations changing the response of a GET} - Sequences generated by this behaviour always start and end with the same GET operation. Thus, there will be one exploration for every available GET operation in the OAS. By always starting and ending with a GET operation, changes in the state of the SUT can be detected. Between the GETs, there is a sequence of potential state-changing operations. When generating parameters for these operations, in addition to generating random values, an operation can reuse parameters from a previous operation in the sequence. In this way, we can perform multiple operations on the same entity, or use a previous entity as an argument to another operation, reaching further into the state of the SUT. Figure~\ref{fig:state-mutation-example} shows an example generated by this behaviour. The sequence generated in the example, considering the HTTP methods, has the structure of GET-POST-GET. The GET-operations are the same, ``getAllProducts'', and the POST-operation in between, ``addProduct'', causes the responses of the first and last invocation of the GET-operations to differ. Figure~\ref{fig:B3-deeper-example} shows an example where an operation, ``addConfiguration'', uses an entity created earlier in the sequence of the example. The B3 behaviour has the ability to produce examples of typical ``Read'' and ``Create'' behaviour, as shown in the generated example. 
    \item \textbf{B4 - A sequence of state-changing operations NOT changing the response of a GET} - This behaviour aims to generate examples where first a change is induced on the state of the SUT, which is then ``cancelled out'', typically by DELETE operations. Such an example would capture both ``create'' and ``delete'' behaviour in relation to a ``read'' operation. As with B3, we need to base the operations around GETs to observe any changes in the state of the SUT. The generator for this behaviour will generate sequences with GET operations before and after any sequence of at least two operations selected from the POST and DELETE operations in the OAS. As was also the case with B3, the relations between operations in the OAS are leveraged in the selection of operations, to avoid uninteresting sequences. 
    
     As hinted in the title of this behaviour, the aim of the check for the behaviour is that we start and end with the same response of the specific GET. A typical example generated from this behaviour would be a GET-POST-DELETE-GET sequence. This means that the state was first read, an entity was created, the same entity was deleted, and we are back getting an equal response as we started with. Figure~\ref{fig:B4-example} shows an example from one of the case studies of this behaviour. The example includes a GET operation between the POST and DELETE, to inform the user of the state change. This GET could optionally be removed based on the intent of communication of the example. This would fit well into a parsing layer for different presentation options and is orthogonal to the method of producing the example.
\end{itemize}

The original QuickREST sought to find examples of operations and inputs which produced 500-status code results, i.e., crashes~\cite{Karlsson-QuickREST-Property-based-Test-Generation-of-OpenAPI-Described-RESTful-APIs-2020}. We have included this ability as a ``fuzzing'' behaviour, in case a user wants to perform fuzzing instead of searching for the defined REST API behaviours.

In this section, we introduced the specific behaviours defined as the first set of behaviours to use in example generation for \rests{}. This is not an exhaustive list by any means, and the general approach presented is open to extensions to other behaviours in accordance with Definition~\ref{def:rest-property}, by defining other generators and checks for behaviours.

\begin{figure}[h]
\begin{minted}[frame=single,linenos, xleftmargin=8pt, framesep=1mm, fontsize=\scriptsize, numbersep=2pt]{clojure}
{:operation-sequence
   [{:operation "getConfigurationsForProduct",
     :parameters [{:name "productName", 
     :value {"productName" "0"}}]}
    {:operation "addProduct",
     :parameters [{:name "productName", 
     :value {"productName" "0"}}]}
    {:operation "addConfiguration",
     :parameters
     [{:name "productName", :value {"productName" "0"}}
      {:name "configurationName", 
      :value {"configurationName" "0"}}]}
    {:operation "getConfigurationsForProduct",
     :parameters
     [{:name "productName", :value {"productName" "0"}}]}
     ]}
\end{minted}
\caption{Example generated by behaviour B3 with multiple operations on the same entity. A configuration is added to an added product in ``features-service'' case-study.}
\label{fig:B3-deeper-example}
\end{figure}

\begin{figure}[h]
\begin{minted}[frame=single,linenos, xleftmargin=8pt, framesep=1mm, fontsize=\scriptsize, numbersep=2pt]{clojure}
{"getProductByName"
  {:operation-sequence
   [{:operation "getProductByName",
     :parameters [{:name "productName", 
     :value {"productName" "B"}}]}
    {:operation "addProduct",
     :parameters [{:name "productName", 
     :value {"productName" "B"}}]}
    {:operation "getProductByName",
     :parameters [{:name "productName", 
     :value {"productName" "B"}}]}
    {:operation "deleteProductByName",
     :parameters [{:name "productName", 
     :value {"productName" "B"}}]}
    {:operation "getProductByName",
     :parameters
     [{:name "productName", 
     :value {"productName" "B"}}]}]}}
\end{minted}
\caption{Example generated by behaviour B4 in ``features-service'' case-study.}
\label{fig:B4-example}
\end{figure}

\subsection{Type-based Relation Finding}\label{sec:relation-finding}

One of the inputs to our method is an OpenAPI specification (OAS). The OAS describes the available API operations, their inputs and outputs, and any potential types used. However, the OAS does not specify relations and constraints between the operations and their inputs and outputs. This fact results in a common challenge of \rests{} test generation, finding dependencies among operations~\cite{Kim-Automated-Test-Generation-for-REST-APIs-No-Time-to-Rest-Yet-2022}. Solving this challenge is required to generate operations in meaningful sequences, where one operation might depend on another.

When exploring different sequences of operations, in a stateful API, relations between operations are relevant. Such relations might be in the order the operations are executed, and also between the parameters and responses of the operations. For example, a value from a response of one operation might be used as the parameter of another operation. Given the ability to find relations, we are able to generate sequences of operations that perform operations on the same SUT entities. As an example, if we first create a person with operation \textit{A}, we might want to refer back to the same person when we delete the entity in operation \textit{B}. Hence, operations A and B are related via their parameters.

\begin{figure*}
    \centering
    \includegraphics[width=\textwidth]{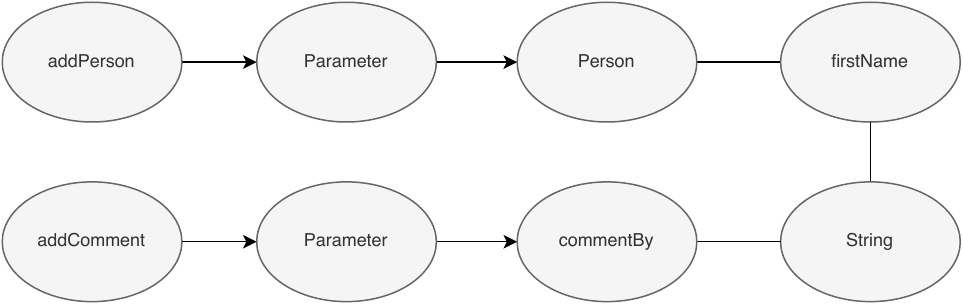}
    \caption{Example of a type-based graph relation. Note the use of both directed and bidirectional relations.}
    \label{fig:graph-type-relations}
\end{figure*}

In this paper, we use a graph-based approach to find relations, similar to what other test generation approaches do~\cite{Viglianisi-RESTTESTGEN-Automated-Black-Box-Testing-of-RESTful-APIs-2020, Zhang-Resource-and-dependency-based-test-case-generation-for-RESTful-Web-services-2021}. 
However, we base our relation graph on the \emph{types} of the operation parameters and responses in the OAS. Building a graph of the relations has been done before, but the relations have included the names of operations and parameters, and their URL resource relations \cite{Viglianisi-RESTTESTGEN-Automated-Black-Box-Testing-of-RESTful-APIs-2020, Zhang-Resource-and-dependency-based-test-case-generation-for-RESTful-Web-services-2021}. We only consider the types, down to scalar types, as, for example, strings. By building a type-based graph we can find relations that are not resource or name related, i.e., the operations might not share a common URL in the OAS. For example, imagine an API operation ``addPerson'' which takes as an input a ``Person''-object with a field of ``firstName''. The first-name is marked as a ``string'' in the OAS. Another operation, ``addComment'' has a parameter called ``commentBy'' of type string. The operations do not share anything in the OAS, one has a parameter of ``Person'' and one of a plain string. However, with a type-based graph, there will be a path between these operations parameters, via the common string type. Hence, we can find the relationship when generating examples, between the ``firstName'' of a ``Person'' and the ``commentBy'' parameter.
The example show us that there is a type-based connection between the ``addPerson'' and ``addComments'' operation, with the path shown in Figure~\ref{fig:graph-type-relations}. However, as with all solutions, there are trade-offs. With this method, compared to URL-based relation-graphs, we can find relations even if there are no shared URL or names in the operations, parameters and responses. The potential downside is that the graph search-space is larger since the graph is on a lower level, i.e., containing more nodes and relations. In the context of generating examples, the solution to the relation problem is just a means to be able to generate good examples. Thus, we do not make any claims or evaluate how this approach compare to other proposed relation finding approaches.

How the OAS is structured plays a part in how large the search-space will be. Two different ways of defining the types of parameters and responses in an OAS are either in-line, together with the operation, or as a reference to a type defined in the ``Definitions'' part of the specification. For example, we can in the same part as we state that an operation takes a parameter state that it consists of an ``age'' of int and a ``name'' of string, or make a reference to a ``Person'' type with the same fields, expressed in the definitions of the specification. Thus, we get two different versions of types; nominal (by name) or structural (by similar structure). The nominal case means that all operations that use a reference to ``Person'' mean the same thing, whereas if two operations locally define the same structure of a type, we can not be sure it is the same thing. For example, two local definitions of ``age'' and ``name'' could both be referring to persons, i.e. related, or one is referring to cars, i.e. unrelated. Consequently, it is more costly to find potential relations in the structural case than in the nominal case. With named types, we know the type is shared.

We consider the distance when navigating the graph to find relations, i.e., the closer two parameters or responses are, the more likely they are deemed to concern the same domain entity. Each part of a structural type will be a node. Following the previous example, there will be 4 nodes $[firstName:string] \rightarrow [string]$ and $[age:int] \rightarrow [int]$. If another parameter shares part of its structure then it will point to $[firstName:string]$, if not it might point to $[string]$ if it has a string parameter. The first case is thus deemed as a closer relation than the second case. Two parameters with relation to the same nominal or structural type will thus be more closely related than being related via a scalar type.

In summary, from the types defined in the given OAS, we create a graph based on these types. The graph also connects the operations, parameters, and responses to these types. We support both types declared in the OAS as nominal, i.e., references to named types, or structural, i.e., inlined definitions in the response/parameter declaration. Leveraging the paths found in this graph, we can test for probable relations when generating example trials.

\subsection{Shrinking Examples}

To maximize the information density in the generated examples presented to a user, they should be minimal. The example generation process generates trial sequences and checks those for conformance of the sought behaviour, as described. Such a generated example might conform to a behaviour, but might still not be \emph{minimal}. A minimal example would be an example only including the required operations to achieve the behaviour. For example, the generated example in Figure~\ref{fig:state-mutation-example} shows a GET-POST-GET sequence and is a minimal example to show the behaviour of creating an entity that affects the specific GETs. However, the same conformance to the behaviour would be reached if an arbitrary number of additional POSTs were inserted in the sequence. Inserting such redundant operations is not helpful to a user, it only adds noise and makes the example harder to read.

To generate minimal examples, we use \emph{shrinking} of any trial examples that conform to a behaviour. The expanded tool, QuickREST, is based on property-based testing~\cite{Karlsson-QuickREST-Property-based-Test-Generation-of-OpenAPI-Described-RESTful-APIs-2020}. Shrinking, i.e., producing minimal failing examples is a typical part of PBT~\cite{Claessen-QuickCheck-2000}. However, as with generators, the included shrinking algorithms are not domain aware and too naive for our use-case. The shrinking algorithm we use in our method tries to shrink the parameter values used when executing an operation, and also tries to shrink the sequence of operations. In the shrinking of the sequence our implementation respect relations. This means that if an operation refers to a value from an operation prior to the sequence, that dependency is not shrunk away. Dependent operations must all be removed or included in a trial of a shorter sequence. This strategy avoids shrinking trials with examples with broken references among operations. For example, in a sequence of GET-POST(x)-GET-GET-DELETE(x)-GET, we can shrink away the redundant GETs and the DELETE---no operation has a dependency to DELETE later in the sequence. However, we do not shrink away the POST(x), since the DELETE operation has a reference, and operates on the same value, ``x''. This approach to shrinking generated examples---respecting relations---is similar to the approach proposed in the general example generation approach, proposed by Karlsson et al.~\cite{Karlsson-Exploring-API-Behaviours-Through-Generated-Examples-2023}. But as with other parts of our proposed approach, we instantiate the general approach with the specialised knowledge of dealing with REST APIs. For example, we do not want to shrink away the GET-operations for behaviours which depend on the return value of the GETs to evaluate the conformance of the example.

Another challenge in providing minimal examples is the continuous change of the state in the system, such as entities being added and deleted in a database. In our approach, we do not expect or require the user to provide any reset functionality. The problem regarding state and shrinking is that, due to changes in state, an example that failed a behaviour conformance check might succeed in the next shrinking trial, and vice versa. The consequence of this is that our shrinking algorithm can not always produce the smallest example, some ``noise operations'' can be left, due to the SUT state changing during the shrinking process. The positive side of this trade-off is that resetting the state in this kind of system-level testing, as \rest{} testing often is, is costly, a cost we do not need to pay, for each trial.

\subsection{The Resulting Artefact}

\begin{figure*}
    \centering
    \includegraphics[width=\textwidth]{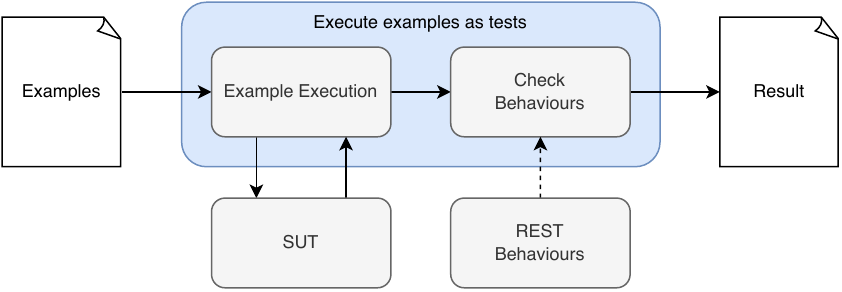}
    \caption{Overview of the execution of examples as tests}
    \label{fig:overview-execution}
\end{figure*}

The resulting artefact of running a test-generation or fuzzing tool is important to consider. The process of searching for test cases can be a time and resource-consuming process, while on the other hand executing the actual resulting artefact, such as a test suite, can be considerably less resource-consuming. As an example, the SotA method EvoMaster is recommended to run between 1 and 24 hours while searching for a test suite with high coverage and fault finding~\cite{Evomaster-url}. The result is a test suite, in a configurable output format such as JUnit-format. Running the resulting suite can typically be executed in seconds or minutes, a large contrast to the time it might have taken to generate the suite. In addition, the resulting suite can be used in regression testing, complemented by humans, stored in source control, etc. Reasons such as these make it important for test generation tools to have an executable output, which several of the current SotA tools lack~\cite{Zhang-Open-Problems-in-Fuzzing-REST-2022}.

However, there are downsides to producing a generated test suite in a test-framework format, such as, for example, JUnit. To execute such tests you now have a dependency on the test-framework used. As a user, you need to set up such a project to be able to run the artefact. This potential problem is amplified in a black-box scenario, where the technology of the SUT is not shared with the test suite, for example, the SUT might be written in PHP while the generated tests are in Java. You now need to understand how to execute tests expressed in another technology. Choosing to output the tests in the format of a common framework might make it easy to execute them---if the organization is familiar with the technology---and since the output then must be formatted as source code, there is a risk of readability issues. 

The last point we want to highlight is also related to a source-code-based artefact. If instead of producing source-code, an approach provides its output in a common and well-defined data-structure format such as JSON~\cite{JSON-url}, it will allow users to further use the output.
For example, if the user would like to produce an HTML-page with documentation based on the outcomes of the test-generation, making such programs is simplified if the test-generation artefact is straightforward to parse.

We acknowledge the importance of an executable artefact, as enumerated above, and also try to address the other mentioned issues. Our extension to QuickREST, which generates examples of behavioural properties, is able to execute its own output. As long as you have the tool, you can execute the generated tests without other dependencies. This is enabled by using a sub-part of the example generation process---as is shown in Figure~\ref{fig:overview-execution}, instead of generating an example candidate, we execute and check an existing found example. The format of the output is a list of the operations in the example and any parameters, as in Figure~\ref{fig:state-mutation-example}. The example is concise, helping readability, and regular, helping any further processing. Further studies are needed to evaluate which kind of output developers and testers prefer and in which scenarios. However, readability has been identified as an important challenge in REST API test generation \cite{Zhang-Fuzzing-Microservices-In-Industry-2022}, and with our method, we provide an alternative to the test-framework-based approach. Figure~\ref{fig:state-identity-example} shows a longer generated example where an entity, a ``product'' in this case, is both created and deleted. Both examples in Figure ~\ref{fig:state-mutation-example} and Figure ~\ref{fig:state-identity-example} are output from our case-study on the ``feature-service'' SUT.

\begin{figure}
 \begin{minted}[frame=single,linenos, xleftmargin=8pt, framesep=1mm, fontsize=\scriptsize, numbersep=2pt]{clojure}
{"getAllProducts"
  {:operation-sequence
   [{:operation "getAllProducts", :parameters []}
    {:operation "addProduct",
     :parameters [{:name "productName", 
                   :value {"productName" "0"}}]}
    {:operation "getAllProducts", :parameters []}]}}
\end{minted}
    
    \caption{Generated state-changing example}
    \label{fig:state-mutation-example}
\end{figure}

\begin{figure}[h]
\begin{minted}[frame=single,linenos, xleftmargin=8pt, framesep=1mm, fontsize=\scriptsize, numbersep=2pt]{clojure}
{"getAllProducts"
  {:operation-sequence
   [{:operation "getAllProducts", :parameters []}
    {:operation "addProduct",
     :parameters [{:name "productName", 
                   :value {"productName" "A"}}]}
    {:operation "getAllProducts", :parameters []}
    {:operation "deleteProductByName",
     :parameters [{:name "productName", 
                   :value {"productName" "A"}}]}
    {:operation "getAllProducts", :parameters []}]}}
\end{minted}
\caption{Generated example of creation and delete behaviour}
\label{fig:state-identity-example}
\end{figure}

\section{Evaluation: Relevance}\label{sec:evaluation-relevance}

The goal of the approach presented in this paper is to generate examples that can be used to advance the understanding of the SUT and to be used as test cases. We evaluate these two different aspects separately. In this section we present the evaluation with regards to generating \emph{relevant} examples.

In order to evaluate if the examples generated by our proposed approach are \emph{relevant}, we have conducted focus group sessions with industry practitioners. Focus groups are valuable to get early feedback on an approach, before investing further in the idea~\cite{Kontio-Using-the-focus-group-method-in-software-engineering-obtaining-practitioner-and-user-experiences-2004}. In addition, the openness of the evaluation method, provides an opportunity to capture a broad spectrum of feedback from practitioners~\cite{Kontio-Using-the-focus-group-method-in-software-engineering-obtaining-practitioner-and-user-experiences-2004}.

In this section we evaluate the research question of; \textbf{RQ1}: How do practitioners perceive the relevance of the generated examples?

\subsection{Focus Group and Questionnaire Setup}

\begin{figure}
    \centering
    \includegraphics[width=.7\textwidth]{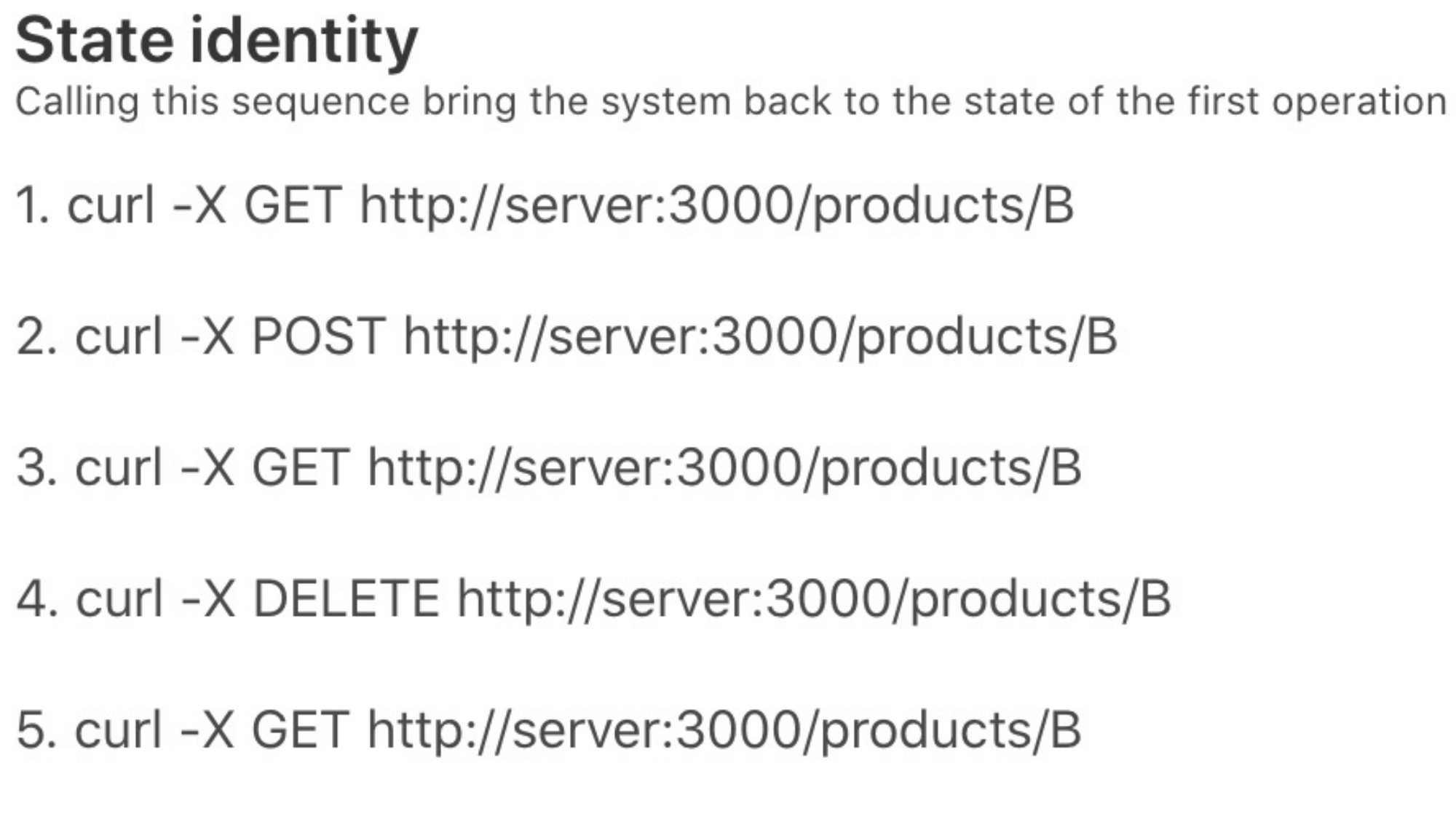}
    \caption{A generated example shown to the participants during the focus group session.}
    \label{fig:report-example-state-identity}
\end{figure}

\begin{table*}[t]
    \centering
    \caption{Quantitative data from the questioner. The scale of the questions is: 1-Very low, 2-Low, 3-Moderate, 4-High, 5-Very High}
    \begin{tabular}{c|l|c}
    \hline
        Id & Question & Avg. \\
    \hline
         Q1 & \textit{``I would rate my experience with REST APIs as''} &  2.9 \\
         \hline
         Q2 & \textit{``How would you like to rate the readability of the generated examples?''} & 3.5 \\
         \hline
         Q3 & \textit{``How would you like to rate the quality of the generated examples?''} & 4.0\\
         \hline
         Q4 & \textit{``How likely is it that you would keep the generated examples?''} & 3.8 \\
         \hline
         Q5 & \textit{``To what degree do you think an example generation method could} & \\
          & \textit{ help you better understand an API that you would use?''} & 4.2\\
         \hline
         Q6 & \textit{``To what degree do you think an example generation method could } &  \\
          & \textit{ help you better understand an API that you develop?''} & 4.3 \\
         \hline
         Q7 & \textit{``To what degree do you think an example generation method could } & \\
          & \textit{help you better test an API?''} & 4.6\\
         \hline
         Q8 & \textit{``To what degree do you think an example generation method could} & \\
          & \textit{help you better document an API?''} & 3.8 \\
    \hline
    \end{tabular}
    \label{tab:questioner-quantitative}
\end{table*}

In order to understand if practitioners deem the generated examples as relevant and helpful, we performed two focus group sessions with two different agile teams at our industry partner. The two teams are development teams producing features for a digitalization platform for factory automation systems. REST APIs are one of the main types of APIs used in this product.

15 practitioners in total participated in the sessions, including the two teams with the addition of some cross-team roles. The largest role group in the sessions was Software Engineers (10). The groups also included Product Owners (2), Quality Assurance Engineers (2), and a DevOps engineer (1). Performing the sessions with two complete agile teams---including supporting roles---gave us diversity in roles and experiences. The number of years of professional experience was also diverse among the participants, ranging from 2 to 24 years with an average of 13.2 years, based on questionnaire data.

The sessions were conducted as online Microsoft Teams meetings. Two of the authors of this paper were present. The first author of this paper took the role of the main moderator, driving the discussion forward. The second author as an observer and complementary moderator, posing further follow-up questions from the discussions.

The procedure of the sessions, inspired by the practical guidelines in~\cite{Breen-A-Practical-Guide-to-Focus-Group-Research-2006}, was as follows; the main moderator introduced the participants to the agenda, the proposed approach of example generation for REST APIs, and the goal of the session---to collect the participants thoughts about the generated examples. Approval for recording the session was obtained. After an introduction to the topic, examples from all the proposed behavioural properties from Section~\ref{sec:rest-behaviour-properties} were discussed. The examples were generated by our proposed approach executed on the ``features-service'', one of the services used in other studies~\cite{Zhang-Open-Problems-in-Fuzzing-REST-2022, Kim-Automated-Test-Generation-for-REST-APIs-No-Time-to-Rest-Yet-2022}. This is a real-world service representative of a RESTful API where high coverage can be achieved, which is important since, if a service cannot be covered, we cannot generate examples for it. The service contains 18 endpoints which makes it not too trivial and not too complex to understand. By generating examples using a service the participants are not familiar with, any understanding of the service gained by the participants is from the generated examples presented in the focus group session and not biased by previous exposure to the service. A generated example discussed by the participants is shown in Figure~\ref{fig:report-example-state-identity}.

In addition to the focus group sessions, we also offered a voluntary questionnaire to perform after the session. This allowed us to capture some quantitative data. 10 participants chose to perform the questionnaire. Table~\ref{tab:questioner-quantitative} shows data from the questionnaire. The data collected from the questionnaire will be discussed and put into context in the presentation of the qualitative results.

\subsection{Results RQ1 - Relevance}

\begin{table*}[t]
    \centering
    \caption{Themes and codes}
    \begin{tabular}{l|l}
    \hline
        Theme & Codes\\
    \hline
         Current State of Practice & Currently only for using other APIs \\
         \cline{2-2}
         & We should make examples for our own APIs \\
         \cline{2-2}
         & Example usages \\
         \cline{2-2}
         & Lack of examples \\
         
         \hline
         Using examples for understanding & Dependency on requirements \\
         \cline{2-2}
         & Combination says more than individual ones \\
         \cline{2-2}
         & ``Debugging'' the systems through the examples \\
         \hline
         Desired improvements & Want to see responses \\
         \cline{2-2}
         & User experience improvements \\
         \cline{2-2}
         & Authorization \\
         \cline{2-2}
         & ``Relations'' not clear \\
         \cline{2-2}
         & More clear state changes \\
         \hline
         These examples are good & Usage of the examples \\
         \cline{2-2}
         & Usage in industrial state of practice \\
         \cline{2-2}
         & Automation is good \\
         \cline{2-2}
         & These examples are helpful to be kept \\
         \cline{2-2}
         & Aids in verification \\
         \cline{2-2}
         & Aids in understanding \\
         \cline{2-2}
         & Aids in test \\
         \cline{2-2}
         & Aids in design \\
         \cline{2-2}
         & REST best practices \\
         \hline
         Examples are good & Examples in general are good \\
         \cline{2-2}
         & Lack of examples is a limitation \\
         \hline
         Expressing examples & Curl is good \\
         \cline{2-2}
         & Swagger/OpenAPI is good \\
         
    \hline
    \end{tabular}
    \label{tab:themes}
\end{table*}

To analyse the result from the focus group sessions we made a thematic analysis of the transcriptions. Quotes from the participants were coded based on the content of the statement. After coding the statements, larger themes were created to capture the essence of the participants' views. An overview of the themes and codes is shown in Table~\ref{tab:themes}.

Some of the main themes and codes identified relate to \emph{understanding and analysis of the system's behaviour} and different \emph{usage scenarios} for the generated examples. In addition, we identified themes touching on different quality aspects of the generated examples and a \emph{current state of practice} theme. We will present the results of the main themes with supporting quotes from the participants that we include throughout the text in \textit{italics}.

\subsubsection{Current State of Practice}

The participants use examples today, primarily to understand APIs they are using: \textit{``The X API has some documentation with a set of examples and I have read through that and still do''}. Usage scenarios also include understanding how to test APIs; \textit{``from a test perspective we do consume all kinds of documentations including examples''}. Examples are seen as very helpful, and not having them is limiting. This is in line with the current literature---examples are very useful~\cite{Robillard-What-Makes-APIs-Hard-to-Learn?-Answers-from-Developers-2009}. As one participant commented, from a testing perspective: \textit{``not having examples is a severe limitation''}.

The engineers also recognised that it would be beneficial to create examples of the APIs they develop; \textit{``It would be helpful both for ourselves and everyone that will use our API''}. Putting themselves in the situation of a user helps the design process; \textit{``producing examples will force you to think through the workflow of a user of the API''}. However, producing examples is currently lacking: \textit{``That's not something we do, but we probably should, because I think it will be helpful''}.

In summary, the engineers see API usage examples as helpful and important, but often do not produce them. Thus, our proposed approach to automatically produce examples of behaviours could fill this gap. However, to do so, the generated examples must be relevant to the engineers, which we discuss next.

\subsubsection{Understanding and Analysing the Behaviour of the System}

A common theme we observed was that the participants would understand some behaviour of the system based on the examples shown; \textit{``I can infer that by posting a new product it will be added to the list of products. So the POST adds a new product''}. Figure~\ref{fig:report-example-state-identity} shows an instance of such an example.
The engineers could draw specific conclusions from the examples; \textit{``This is how I create a new product in the system''}. Having examples also tells more about the behaviour than what you can infer from only looking at the HTTP method (GET, POST, PUT, DELETE) of the operation; \textit{``[Examples] tells a bit more about operations than just  HTTP methods does.''}

The participants gained a deeper understanding of the effects of performing a sequence of operations would have on the state of the system. Thus, this understanding goes beyond just the responses of the operations in the example; \textit{``This tells me that this particular POST operation actually does. [change the state]''}, and another engineer noted; \textit{``Another kind of observation I think you can draw from this is that Products seem to be able to *really* create and then *really* delete. Meaning that if I do a DELETE it's gone, there is nothing left of the product that I created''}. As the participant explained in relation to the previous quote, in some systems he is working with entities are not actually deleted, only marked as such. In addition, that leads to an analysis of the effect on the state of the system as a consequence of the behaviour displayed by the examples, which the example did not explicitly show; \textit{``I would assume that I could create a product B again. And if it didn't show up as a state identity [B4], then I would assume that I cannot create a product anew. I've created it once and now can't do it again''}.

A common theme we observed was that the participants would understand some behaviour of the system from the examples, while also using the examples as a starting point for further analysis of ``what if'' scenarios. The participants reasoned about the examples seen in the context of how the SUT actually behaves, and in doing so also extended this understanding to what it would mean if an example of the given behaviour was different; \textit{``Nothing else is affected by the GET [operation], like a date timestamp or anything like that. But if it did, I would assume that the system under test is *really* stateful.''}, and in addition; \textit{``if we would have an example of a query POST operation that changed the state, that would be very informative.''}

Looking at the behaviour examples in aggregate gave the engineers a deeper understanding of how an operation behaves; \textit{``From both these two examples, the first one was response equality [B1] and we got the example that if you do GET twice, it will not change [the response] and this one [B2] says the same thing but from the other perspective, there are no two GET operations that produce different responses. So these two combined definitely tell me that doing these GET operations reads from the system, nothing outside affects the responses of these GET operations''}. Looking at examples of both B1 and B2 for an operation also provided insights about concepts such as idempotency, in addition to how the operation behaves; \textit{``So the first one [operation invocation] could actually insert or create the product zero and the second one [operation invocation] doesn't do anything because it's already created, it just responds back with the same response as the first one. So from the user perspective of the API, it's idempotent''}.

The generated examples show invocations of at least two operation invocations. This was seen as a benefit and a missing piece of specifications; \textit{``I think this is a great example and I think that it's missing when using API specifications, to see the full flow, the chain of calling methods''}.

The behaviours B1--B4 (defined in Section~\ref{sec:rest-behaviour-properties}) were seen as important, \textit{``For me, these are all equally important. They show some properties of a stateful API that is of relevance. Maybe there are more, but these are really important''}. The participants could not think of any other behaviours in addition to B1--B4 that would be useful in the context of generating examples for RESTful APIs; \textit{``I think that it feels like there isn't anymore, these are the minimized examples that exist, or? Is there any more that doesn't fall into these four categories?''}.

The quantitative data collected with regards to understanding support the qualitative analysis in concluding that the engineers find the generated examples relevant in understanding the APIs they develop (Q6 in Tabel~\ref{tab:questioner-quantitative} with a score of 4.3) and APIs they use (Q5 in Tabel~\ref{tab:questioner-quantitative} with a score of 4.2).

\begin{result}
The participants understand more about how the system works through the generated examples. The generated examples also stimulate further analysis of the effects of the state of the system. Having multiple examples of different behaviours of the system is helpful for the participants in advancing their understanding.
\end{result}

\subsubsection{Usage of the Generated Examples}

In the previous section, we presented the results of the focus group sessions regarding the participants' ability to understand the system under test. In this section, we analyse the results from the theme of \emph{usage}, meaning, in what ways the engineers want to use the generated examples.

In addition to the already discussed usage area of furthering the understanding and triggering of analysis of the behaviour of the system, the participants considered using the examples for verification of behaviours when developing an API, for testing, for documentation, and as an aid when requirements are missing.

The participants thought that the examples would serve as test cases; \textit{``[Example of State Identity] it's a pretty extensive test of the SUT doing all these operations''}, and also as an inspiration for what to test: \textit{``Examples is key to when you don't work with it [the code base] every day and just need to test it somehow''}.

When requirements are missing, generated examples can be a substitute to reason about the behaviour of the system;
\textit{``Examples are really good to drive and understand how the system is supposed to work or at least try to find inconsistencies. As long as we don't have any requirements, that's kind of the only way of doing it. So it's really good to have''}.

In the situation where requirements do exist, the generated examples can be used for verification purposes; \textit{``From [the perspective of] developing an API, we have this specification, it should exhibit this kind of behaviour and it should exhibit that kind of behaviour and we could use this [generated examples] to ensure that we follow the specification, using these abstract properties of stateful APIs''}. In addition, the generated examples can help the engineers to reflect on if they have built what they set out to do; \textit{``The example triggers my brain to look at it and investigate if something looks fishy''}. Having a set of generated examples that cover multiple scenarios provides the engineers with trust in that the system does what it is supposed to do; \textit{``From a general perspective also it adds trust on the system when you have such examples. Many times what happens is you don't have any examples of how the API work or there are partial examples not covering all the scenarios. But if you do have examples of these sorts, it builds your trust''}.

Another usage area, mentioned by the engineers, is documentation. The usage of examples for documentation is relevant both for internal and external purposes; \textit{``When you're designing the API, for internal communication and to have a common understanding of how to give examples and describe to each other and in documentation as well''} and another engineer note that; \textit{``I would use them as documentation to create the regression bed for automating the APIs''}. For external documentation, the generated examples can be a complement, \textit{``I think this [the generated examples] is a good complement to other kinds of documentation''}.

The suggested usage areas in the quantitative survey show that usage in testing scenarios is the strongest area (Q7 in Table~\ref{tab:questioner-quantitative} with a score of 4.6) and that usefulness for documentation is moderate-to-high (scoring 3.8, Q8 in Table~\ref{tab:questioner-quantitative}).

\begin{result}
 Our results show that the examples generated by our proposed approach are useful to the participants in several different areas, such as test cases, verification of developed behaviour and documentation. 
\end{result}

\subsubsection{Additional Quality Aspects of the Generated Examples}

In the overall quantitative quality assessment of the generated examples the participants judged them to be of a high quality (Q3 in Table~\ref{tab:questioner-quantitative}). Quality does not have an exact definition and can mean different things to different participants. In the previous sections, we discussed the aspects of the generated examples helping in understanding the system and how the engineers wanted to use the examples. In this section, we focus on the quality aspects of how the examples are expressed and how the participants wanted to improve them.

The readability of the examples got a quantitative evaluation of 3.5 (Q2 in Table~\ref{tab:questioner-quantitative}), and the qualitative data from the focus group give insights into both the positive aspects of how the examples are expressed and how they can be improved. The examples, as can be seen in Figure~\ref{fig:report-example-state-identity}, were expressed as Curl\cite{Curl} commands. This was seen as very positive: \textit{``I think Curl is great. I mean, that's sort of the industry standard of expressing HTTP calls.''} and language agnostic: \textit{``everyone who works with their language of choice, they can understand Curl''}. However, several engineers pointed out that the examples should include the status code returned and the actual responses from the executions. Our approach has all this data, but it was not included in the report since it can be quite large and make the report harder to read. However, the engineers would like to have the opportunity to browse this data: \textit{``I would still like to have an opportunity to see the entirety of the responses. But I don't think it should be shown right away, it should be an option to show it.''} and also suggested that the differences between the operation executions are shown \textit{``If it would be possible to sort of easily highlight things that change between different responses. For instance, if you have response inequality, what changed? We could show that.''}.

A theme that could explain that the usage as documentation has the score 3.8 (lowest of the usage related questions Q5-Q8) and that also might affect the readability of the examples are the values of parameters used in the examples. If the examples would have been produced manually, the engineers would have included parameter values that better fit the domain of the data, \textit{``I think the manual [examples] we've written have better data, the input and output data. We will get at least a little better user-friendly input and output data.''}.

One suggested improvement that could potentially increase both the understanding of examples and the readability, is to surround potential state-changing operations with GET-operations. As the participants pointed out, for longer generated examples that include multiple state-changing operations (POSTs for example) it is not always clear which of the operations actually changed the state: \textit{``In the last two examples there are 2 POST operations between the two GETs. It's really not clear what POST operation changed the state. If we take the first example into consideration, the 1st POST operation seems to change the state, but do we actually know that the second one does?''} and to fix this: \textit{``I mean, if there would be a GET between all of the POST then you can [know]''}.

\begin{result}
The participants judge the quality of the examples to be high. Curl commands are a good way to express REST API examples. To increase readability and understanding, the generated examples should include responses, domain-related values, and more GET operations.
\end{result}

\begin{result}
\textbf{RQ1}: Based on the evidence presented from the focus groups and survey, we conclude that our proposed approach can generate examples that are \emph{relevant} to engineers; the generated examples help in further the understanding of the API, and is useful for several purposes such as tests and documentation. The suggested improvements are not improvements on the approach, but rather engineering and user-experience work.   
\end{result}

\section{Evaluation: Test Generation}\label{sec:evaluation-test-gen}

In this section we evaluate the test generation part of the proposed approach. We do so by evaluating the following research questions;
\begin{itemize}
    \item \textbf{RQ2}: How does our behaviour-driven test-generation method compare to test-generation with EvoMaster, based on search-coverage? 
    \item \textbf{RQ3}: How do search and execution coverage compare between the methods and what are any limiting factors?
\end{itemize}

In this paper, we present a novel method of generating tests for an OpenAPI-described \rest{} by generating examples of how the SUT behaves. This is in contrast to methods focusing on code-coverage and fault-finding. Even though our method is not focused on code-coverage, but rather covering behaviours, we can still use code-coverage as a proxy measure of how much of the behaviour of the SUT we can reach. Therefore, we evaluate our method based on code-coverage, which is the typical measure used in REST-API method evaluations \cite{Zhang-Open-Problems-in-Fuzzing-REST-2022, Kim-Automated-Test-Generation-for-REST-APIs-No-Time-to-Rest-Yet-2022}. In addition, we want to support the claim that using an example generating approach can be done without giving up test-generation possibilities---in addition to producing a source of better understanding the SUT. Since our focus is on behaviour examples, closer to the business logic of the SUT, we can not judge if a specific example is correct or incorrect. The correctness must be assessed based on the requirements of the SUT. However, once an example is deemed as nominal, it can be used for automatic regression testing. For example, if we find an example of where entities with the same ``name'' property can be created, we can not judge if this is correct or not, only the requirements can tell us. In light of this, we do not measure any fault-finding in our evaluation.

As a comparison with SotA in \rest{} fuzzing, we compare against EvoMaster. The reason to select EvoMaster is that it has been found in multiple recent studies to be the best on average performing method for \rest{} fuzzing \cite{Zhang-Open-Problems-in-Fuzzing-REST-2022, Kim-Automated-Test-Generation-for-REST-APIs-No-Time-to-Rest-Yet-2022}. As mentioned, the goals of EvoMaster and our method are different. Thus the comparison does not aim to evaluate which method is ``better'', but rather to show the potential areas where using a behaviour-based approach can be advantageous.

Since our approach aims at producing examples of behaviours, the outcomes must be thought of in a different way than an approach aiming to generate a test-suite with the goal of code-coverage. Since our approach searches for behaviours, it can also be used to search for examples of behaviours that should not be present. If the intent of the requirements, for example, is that usernames must be unique, the list of examples of that we can POST users with the same username should be empty. Hence, executing the artefact would yield zero coverage of the SUT, which is the expected and correct outcome in this case. On the other hand, if a counter-example is found, indicating a bug, the coverage would increase. Thus, we could see a search-coverage of 100\%, while the execution-coverage of the (empty) test-suite would be 0\%. This example also highlights that what we propose is not an either/or scenario of using our approach instead of a SotA \rest{} fuzzer, but rather as a complement, to assess the behaviours of the SUT, and while doing so, generating usage-examples parsable into documentation formats. 

This evaluation has several goals. The method needs to cover the code of the SUT to have the potential to assess the behaviours of the SUT. We want to know how much of the SUT the behaviour search can reach (RQ2). Secondly, when behaviours are found, it should be possible to consistently execute the generated test-suite (RQ3). Since we do not use any reset of the SUT during the search or the execution of the resulting test cases, we want to learn what the effect of this strategy is on the execution of the test cases. Creating state-independent test-cases are preferred for successful usage in regression cases.

\begin{table*}[t]
    \centering
    \caption{Achieved code-coverage of the evaluated methods. We report the average line coverage, the [min, max] out of 30 searches (S). In addition, the Mann-Whitney-Wilcoxon U-Test p-values (all being $<0.001$) and effect sizes $\hat{A}_{12}$ of EvoMaster vs. QuickREST are included.}
    \begin{tabular}{ l | r | r r c r }
    \hline
         SUT & Base & EvoMaster BB (S) & QuickREST (S) & $\hat{A}_{12}$ S & Time(s) \\
    \hline
         features-service & 18.60 & 57.99 [57.99,57.99] & \textbf{60.33 [56.46,73.96]} & 0.23 & 102.8  \\
         languagetool & 1.32 & \textbf{2.98 [2.47, 11.87]} & 2.47 [ 2.46, 2.47] & 0.76 & 13.6 \\
         ocvn-rest & 10.07 & 25.06 [22.56,26.33] & \textbf{27.48 [27.33,27.49]} & 0.00 & 241.6 \\
         proxyprint & 4.16 & \textbf{30.86 [30.80,31.37]} & 28.41 [28.36,28.50] & 1.00 & 311.0 \\
         rest-ncs & 4.36 & 63.62 [62.91,63.64] & \textbf{89.32 [80.36,93.82]} & 0.00 & 25.8 \\
         rest-news & 12.50 & \textbf{68.06 [68.06,68.06]} & 32.64 [32.64,32.64] & 1.00 & 42.4 \\
         rest-scs & 4.06 & \textbf{62.81 [62.03,63.39]} & 61.44 [61.02,62.03] & 0.99 & 27.5 \\
         scout-api & 12.23 & \textbf{18.55 [18.26,18.86]}& 17.72 [17.25,18.11] & 1.00 & 48.2 \\
    \hline
    \end{tabular}
    \label{tab:evo-qr-coverage-search}
\end{table*}

\subsection{Experimental Setup}

We used EvoMaster version 1.4~\cite{Evomaster-1.4}. We used the documented recommended defaults for black-box testing~\cite{Evomaster-Blackbox-doc}, with one exception. The default suggests using a rate-limit on requests to the SUT with a maximum of 60 per minute to not cause a denial of service attack. However, since QuickREST has no such rate limit, and to be fair when comparing the outcomes, we removed this limit for EvoMaster. Note that removing this rate limit increases the coverage for EvoMaster, thus the configuration of such a setting would be important to state for any evaluations comparing \rest{} tools, as it can affect the results between the tools.

The implementation of our method is based on the original QuickREST implementation~\cite{quickrest-old-url}. In addition to implementing our described approach, we did some engineering additions, such as being able to run the tool as a command-line application, to simplify the execution of the case-studies. Our implementation is openly available~\cite{quick-rest-repository}.

The studies comparing a larger set of \rest{} fuzzers, execute the tools from 10 minutes and up to 24 hours\cite{Kim-Automated-Test-Generation-for-REST-APIs-No-Time-to-Rest-Yet-2022} or for one hour~\cite{Zhang-Open-Problems-in-Fuzzing-REST-2022}. The recommended time for EvoMaster, according to its documentation~\cite{Evomaster-url}, ranges from 1 to 24 hours. 

The time taken for test-generation is of importance in industry~\cite{Zhang-Fuzzing-Microservices-In-Industry-2022}. This is also very intuitive, the faster the user can get a result, the better. There are several considerations to make concerning time. When executing a fuzzing tool on a continuous integration (CI) server it might be fine for practitioners to wait for the result for hours. But \rest{} test-generating tools could also be used to give fast feedback when developing or maintaining functionality, as in a test-driven development approach. For example, in the case of our approach, the user can get a generated example of the faulty behaviour, fix the bug, and verify that the behaviour is not generated on the new version. This kind of interactive development would require \rest{} test-generation tools to be faster, and maybe allow a developer to focus on parts of an API, leaving the deeper fuzzing to the long-running CI-server. The second consideration we do with regard to time is the reset of the SUT. Resetting the SUT with system-level testing, as \rest{} testing is, can be costly~\cite{Zhang-Fuzzing-Microservices-In-Industry-2022}, especially in a black-box scenario where the internal state cannot be reset with a library method. Therefore it would be preferable if these tools are robust to searching and executing tests with a changing SUT state. To evaluate with these points in mind, the evaluation of our method aims at producing a result as fast as possible. The time taken was recorded and the same amount of time was given to EvoMaster. The SUTs were restarted before each new complete execution, not during search or execution of the generated test-suite. 

The selected SUTs were selected from the EvoMaster Benchmark Suite, included in \cite{Zhang-Open-Problems-in-Fuzzing-REST-2022}, that was also included in \cite{Kim-Automated-Test-Generation-for-REST-APIs-No-Time-to-Rest-Yet-2022}. We removed the ``restcountries'' SUT since the original QuickREST lacks the engineering work to process that OAS.

The SUTs and tools were executed on a virtual Ubuntu 20.04 machine, set up with help of the scripts in \cite{Kim-Automated-Test-Generation-for-REST-APIs-No-Time-to-Rest-Yet-2022}. The host for the virtual machine was a MacBook Pro with 2.9Ghz Intel i9 and 16GB RAM. This setup, is typical for what is used at our industry partner, is representative of an environment a developer might use. Coverage was collected with JaCoCo\cite{jacoco-url}. Experiments were executed 30 times for each SUT, considering recommendations evaluating random-based algorithms\cite{Arcuri-A-Hitchhikers-guide-to-statistical-tests-2014}.

\begin{table*}[t]
    \centering
    \caption{Achieved code-coverage of the evaluated methods. We report the average line coverage, the [min, max] out of 30 executions (E). In addition, the Mann-Whitney-Wilcoxon U-Test p-values (all being $<0.001$) and effect sizes $\hat{A}_{12}$ of EvoMaster vs. QuickREST are included.}
    \begin{tabular}{ l | r | r r c }
    \hline
         SUT & Base & EvoMaster BB (E) & QuickREST (E) & $\hat{A}_{12}$ E \\
    \hline
         features-service & 18.60 & 35.23 [35.23,35.23] & \textbf{48.83 [38.07,69.15]} & 0.00  \\
         languagetool & 1.32 & \textbf{1.98 [ 1.91 , 2.40]} & 1.71 [ 1.71 , 1.71] & 1.00 \\
         ocvn-rest & 10.07 & 21.74 [19.69,24.55] & \textbf{25.66 [25.40,25.75]} & 0.00 \\
         proxyprint & 4.16 & 12.26 [05.75,\textbf{28.03}] &  \textbf{27.08} [\textbf{26.64},27.72] & 0.20 \\
         rest-ncs & 4.36 & \textbf{54.13 [52.73,56.36]} & 32.36 [32.36,32.36] & 1.00 \\
         rest-news & 12.50 & \textbf{62.25 [60.42,64.58]} & 26.39 [26.39,26.39] & 1.00 \\
         rest-scs & 4.06 & \textbf{56.70 [53.22,58.64]} & 45.76 [45.76,45.76] & 1.00 \\
         scout-api & 12.23 & \textbf{18.03 [17.55,18.29]} & 12.23 [12.23,12.23] & 1.00 \\
    \hline
    \end{tabular}
    \label{tab:evo-qr-coverage-execution}
\end{table*}

\subsection{Results RQ2 - Search Coverage}

Table~\ref{tab:evo-qr-coverage-search} shows the results from the tools search. Concerning line coverage during search, QuickREST performs slightly higher coverage in two out of eight SUTs; ``features-service'' (+2.34) and ``ocvn-rest'' (+2.42), and considerable better in one, ``rest-ncs'' (+25.7). In four of the eight SUTs, QuickREST perform slightly lower coverage; ``languagetool'' (-0.51), ``proxyprint'' (-2.45), ``rest-scs'' (-1.37), and ``scout-api'' (-0.83). In one case, EvoMaster produces a considerable higher coverage, ``rest-news'' (+35.42). In summary, QuickREST, with a behaviour-based method, results in a similar search coverage in six out of eight SUTs, compared to EvoMaster. The methods have one SUT each with a larger advantage.

Investigating how the search coverage could be increased for each SUT would require an in-depth analysis of the result of each SUT. Such an analysis is out of the scope of this paper. However, we can make some remarks. The ``languagetool'' case is a known hard case for many REST-API fuzzers \cite{Zhang-Open-Problems-in-Fuzzing-REST-2022,Kim-Automated-Test-Generation-for-REST-APIs-No-Time-to-Rest-Yet-2022}. This SUT is not a very \restful{}, as there is no manipulation of resources. The API only exposes two endpoints, where the functionality of the API is to spell-check a piece of text. Such a non-\restful{} is out of the scope of the proposed properties, we specifically target CRUD-based \restfuls{}, which show in the results.

The second point is the low search coverage, compared to EvoMaster, in the ``rest-news'' system. An analysis of the search shows that due to engineering deficiency in QuickREST, the tool does not correctly format the input data to POST (creation) endpoints in this API. The result is ``Unsupported Media Type'' responses (415), which then yield no deeper coverage on these endpoints.

\begin{result}
\textbf{RQ2}: Producing a result as fast as possible, the proposed behaviour-driven test-generation method achieves a search coverage of on average 39.98\%. This is comparable to the state-of-the-art fuzzing tool EvoMaster with an on-average search coverage of 41.24\%, given the same amount of time.
\end{result}

\subsection{Results RQ3 - Execution Coverage}

Table \ref{tab:evo-qr-coverage-execution} shows the execution coverage of the two approaches, with the min and max included. If the goal of an approach is to produce a test-suite with high coverage, then the search and execution coverage should be as close to each other as possible. However, for QuickREST, as mentioned, this is not the goal. Our approach aims at producing examples of behaviours the user might expect, or be surprised by. Hence, it is perfectly fine for the approach to produce a high search coverage, but a low execution coverage, when we do not expect any of the behaviours searched for. In light of this, how do we then evaluate the execution coverage of QuickREST? Even if we do not know the ideal execution coverage for a set of behaviours---without a deep analysis of the actual behaviour of the SUT---we do at least know that the result for the same set of properties on the same SUT should be \emph{consistent}.

QuickREST has a high degree of consistency in seven out of eight of the SUTs, with a small or zero difference in the min and max. The result for one of the SUTs, ``features-service'', shows a high difference in the average coverage and the min/max results (-10.76, 20.32). This means that for this SUT it is possible to find behaviours with an execution coverage of 69.15\% (the max result from Table \ref{tab:evo-qr-coverage-execution}). A reasonable expectation is thus that the approach should find those on each execution. We can also note that this same SUT, is the one that EvoMaster show the highest difference between search and execution coverage  (22.76\%). 
For QuickREST the major problem with this SUT, is that examples of the same behaviours are not consistently found. Improvement in the search for behaviours is needed to increase the consistency, while at the same time not significantly increasing the time spent in search. EvoMaster shows a different problem. While EvoMaster is very consistent in both the search and execution coverage for ``features-service'', it produces test-cases that are dependent on the exact state of the SUT as it was when searching. Hence, when these tests are executed after a reset of the SUT, or at a later time with a different state, the test fails. For example, a test tries to delete an entity with a specific id, without making sure such an entity is first created. Such a test fails when the entity is not present, thus the execution coverage is lower.

\begin{result}
\textbf{RQ3}: The main limiting factor for QuickRESTs behaviour-driven method to increase the consistency in execution coverage is to decrease the variability in the search result of the SUT's behaviour.
\end{result}

\section{Discussion}\label{sec:discussion}

\begin{figure}
\begin{minted}[frame=single,linenos, xleftmargin=8pt, framesep=1mm, fontsize=\scriptsize, numbersep=2pt]{java}
@Test
public void test_1() throws Exception {
  given().accept("application/json")
         .get(baseUrlOfSut + "/products")
         .then()
         .statusCode(200)
         .assertThat()
         .contentType("application/json")
         .body("size()", equalTo(5))
         .body("", hasItems("ELEARNING_SITE", 
         "NRIuqSIDyR", "H7uPs1", "YPofE4WoBC", "nJojT"));
}
\end{minted}
\caption{State-dependant ``GET'' test-case generated by EvoMaster}
\label{fig:evo-state-dependant-get}
\end{figure}

\begin{figure}
\begin{minted}[frame=single,linenos, xleftmargin=8pt, framesep=1mm, fontsize=\scriptsize, numbersep=2pt]{java}
@Test
public void test_6() throws Exception {
  given().accept("*/*")
         .delete(baseUrlOfSut + "/products/7")
         .then()
         .statusCode(204)
         .assertThat()
         .body(isEmptyOrNullString());
    }
}
\end{minted}
\caption{State-dependant ``DELETE'' test-case generated by EvoMaster}
\label{fig:evo-state-dependant-delete}
\end{figure}

While researchers have made great improvements in the code covered and the faults found with REST API fuzzing, there are plenty of open challenges. Challenges include; handling state in the SUT~\cite{Zhang-Fuzzing-Microservices-In-Industry-2022}, using testing criteria more related to business logic~\cite{Zhang-Fuzzing-Microservices-In-Industry-2022}, the time taken while searching for test cases~\cite{Zhang-Fuzzing-Microservices-In-Industry-2022}, the readability of generated tests~\cite{Zhang-Fuzzing-Microservices-In-Industry-2022}, the optimization of computational resources~\cite{Martin-Lopez-Testing-of-RESTful-APIs-Promises-and-Challenges-2022}, stronger support for stateful testing~\cite{Kim-Automated-Test-Generation-for-REST-APIs-No-Time-to-Rest-Yet-2022}, and finding dependencies between operations~\cite{Kim-Automated-Test-Generation-for-REST-APIs-No-Time-to-Rest-Yet-2022}. In addition, several state-of-the-art methods do not produce an executable artefact when the fuzzing process is completed~\cite{Zhang-Open-Problems-in-Fuzzing-REST-2022}. This greatly reduces the possibility of fast regression testing for practitioners, as re-doing the complete search/fuzzing process is wasteful. Also, we argue, there is a missed opportunity in the search process; to improve the understanding of the SUT, and how the SUT relates to certain \emph{behaviours}.

With the evaluations of the proposed approach, we have shown that an example-generation approach can address several of the open challenges. The generated examples are deemed relevant by practitioners, increasing their understanding of the SUT. In addition, the practitioners find several different uses for the generated examples, such as a source of documentation and verification of best practices. Generating an artefact with multiple usage areas means that an example-generating approach has the potential to provide a higher yield on the time spent searching for test cases. In addition, the generated examples are on the higher abstraction level of behaviours, compared to test cases generated to further a coverage metric. This brings them closer to the level of abstraction of business logic.

One big difference between our approach compared to approaches that produce an executable artefact with specific values in the test-cases, is that the behaviour-based examples are more state-independent. This is something we could observe with the comparison to EvoMaster.
Examples generated from the ``features-service'' case-study highlight this difference. Consider the example generated by EvoMaster in Figure~\ref{fig:evo-state-dependant-get}. We can see in this example a ``get'' of ``/products'' (L4) and the exact specific items expected (L10-11). It is very unlikely that this test will ever succeed after its generation since it is highly specific. To make this test pass on a restarted SUT, the test-suite would need to include the creation of these entities---which it does not. If we instead look at what our method produces, in Figure \ref{fig:state-mutation-example}, we can observe that the behaviour of the same ``get''-operation is exercised in a more state-independent fashion. It makes no difference in this test how many other entities are in the system, the test asserts the behaviour that the result of the ``get''-operation is affected by the ``post''-operation, which is the essence of the ``get''-operation.
A similar situation can be seen in Figure \ref{fig:evo-state-dependant-delete}. In this test, there is a ``delete'' on the entity ``/products/7'' (L4) which is expected to be successful (L6). However, once again, the success of this test relies on that there exists an entity with id ``7'', which is not created by this test, or the test-suite. In contrast, the behaviour-based method for deletion produces the example in Figure~\ref{fig:state-identity-example}. In this example an entity is first created to then be deleted, putting the state of the ``get'' operation back to where we started. This test is self-contained and highlights a behaviour, it also works on a restarted system.

Tests created as examples of behaviours are thus less dependent on the state of the system since they are expressed on a higher level of abstraction---the exact values of entities are not evaluated, but rather the overall behaviour. While our method currently produces tests that do not depend on the state of the SUT, future work is to make sure this invariant holds for a complete set of behaviours, i.e., values should not be reused over a complete set of behaviours.
If you do want generated tests to be very specific on the values returned, the EvoMaster approach is a good complement---\emph{if} the test-suite would create the required conditions for successful execution.

Another strength of the proposed behaviour-based approach is that the user gets control of the search on a higher abstraction level. The user can choose what behaviours to search for---some behaviours might be irrelevant to the type of SUT the user has---thus time and energy will only be spent towards that goal. All the proposed behaviours in this paper can be individually used. This is in contrast to approaches with a more opaque search method with less high-level control for the user.

The final strength of the proposed approach we highlight is the potential for the generated examples to be used for more than test-cases. As we have shown, the generated examples are self-contained. The examples can be post-processed into other formats since the output is a data-structure---not source code. Behavioural examples have been shown to help users in understanding a system\cite{Gerdes-Understanding-Formal-Specifications-through-Good-Examples-2018}. This is an area that could be explored further, for example, to study how users prefer to have generated usage examples of \restfuls{} presented---based on our focus groups, Curl is well received, but there might be better alternatives not yet investigated.

A weakness of a behaviour-based approach is that if the behaviours do not correspond to the domain of the SUT, they will not produce a good search result. Since the proposed first set of behaviour-properties in this paper are targeted towards \restfuls{}, APIs such as ``language-tool'' that are not very RESTful are not applicable for our approach with the current proposed behaviours. Indeed, the method is only as strong as the match between the domain of the SUT and the behaviours used.

Finally, regarding the usage of the proposed approach. We foresee the approach to be useful in a development process where software engineers can use it in an interactive fashion to understand if the API under development is providing the expected behaviours. The engineer can generate examples, analyze the result, make changes to the software, and generate new examples. When the engineer is satisfied with the examples and the behaviour of the software, the generated examples can be saved and executed on new versions of the software, as regression tests. 
Quality assurance engineers can also benefit from the approach, for example when integrating and understanding APIs created by different engineering teams, or as support when understanding undocumented APIs. 
The usefulness of the approach in these cases is supported by the results of our evaluation of the relevance of the generated examples in Section~\ref{sec:evaluation-relevance}. With these use cases in mind, we can see a combination of approaches as beneficial. The approach proposed in this paper might be applied in the active development cycle of an engineer---make a change, assess the result---, while an approach focusing on code coverage and fault finding, such as EvoMaster, can complement when the change to the software is complete---for example, running on a continuous integration server.

\begin{result}
Our proposed behaviour-based test-generation approach produces test-cases that are on a higher abstraction level. They are state-independent, allow the user to have control of the selection of behaviours, and have the potential for extensions of its output to areas such as documentation. But the method is limited to the relation between the domain and the specific behaviours. In order to increase the search result for non-RESTful APIs, additional behaviours are needed.
\end{result}

\section{Related Work}\label{sec:related-work}

\rest{} fuzzing and test-generation have been a thriving research area in recent years, with many methods proposed\cite{Arcuri-RESTful-API-Automated-Test-Case-Generation-with-EvoMaster-2019, Ed-douibi-Automatic-Generation-of-Test-Cases-for-REST-APIs-A-Specification-Based-Approach-2018, Atlidakis-RESTler-Stateful-REST-API-Fuzzing-2019, Karlsson-QuickREST-Property-based-Test-Generation-of-OpenAPI-Described-RESTful-APIs-2020, Viglianisi-RESTTESTGEN-Automated-Black-Box-Testing-of-RESTful-APIs-2020, Martin-Lopez-RESTest-Automated-Black-Box-Testing-of-RESTful-Web-APIs-2021, Laranjeiro-A-Black-Box-Tool-for-Robustness-Testing-of-REST-Services-2021, Wu-Combinatorial-Testing-of-RESTful-APIs-2022, Atlidakis-Pythia-Grammar-Based-Fuzzing-of-REST-APIs-2020, Atlidakis-Checking-Security-Properties-of-Cloud-Service-REST-APIs-2020, Corradini-Automated-black-box-testing-of-nominal-and-error-scenarios-in-RESTful-APIs-2022, Godefroid-Intelligent-REST-API-Data-Fuzzing-2020, Segura-Metamorphic-Testing-of-RESTful-Web-APIs-2018, Stallenberg-Improving-Test-Case-Generation-for-REST-APIs-Through-Hierarchical-Clustering-2021}. However, the main test oracles used focus on finding crashes and conformance between the SUTs API and the OAS. There are exceptions, such as a focus on security properties\cite{Atlidakis-Checking-Security-Properties-of-Cloud-Service-REST-APIs-2020} or breaking changes between API-versions\cite{Godefroid-Differential-Regression-Testing-for-REST-APIs-2020}. In addition, metamorphic testing has been used to find bugs regarding common \rest{} relations, manually defined\cite{Segura-Metamorphic-Testing-of-RESTful-Web-APIs-2018}. To the best of our knowledge, no other work has focused on common behavioural properties, automatically, related to business logic, as we do.

In the area of \rest{} fuzzing, there are two other proposed methods based on property-based testing\cite{Claessen-QuickCheck-2000}, QuickREST\cite{Karlsson-QuickREST-Property-based-Test-Generation-of-OpenAPI-Described-RESTful-APIs-2020} and Schemathesis\cite{Hatfield-Dodds-Deriving-Semantics-Aware-Fuzzers-from-Web-API-Schemas-X-2021}. The properties in these methods, as in the general case for other methods, focus on finding API crashes, conformance to specification, and in the case of Schemathesis, the possibility for performance checks, and conformance to HTTP semantics (such as content type). As described, our method is not primarily targeting fuzzing, but behaviours expected of a CRUD-based \rests{}.

EvoMaster\cite{Arcuri-RESTful-API-Automated-Test-Case-Generation-with-EvoMaster-2019} have been extended with resource-based templates\cite{Zhang-Resource-and-dependency-based-test-case-generation-for-RESTful-Web-services-2021} to improve in performing multiple operations on the same resource, to achieve higher code coverage of the tests. RESTTESTGEN also base their dependencies on resource relations \cite{Viglianisi-RESTTESTGEN-Automated-Black-Box-Testing-of-RESTful-APIs-2020}. We use a type-based graph, which is less sensitive to APIs not fully following \rest{} guidelines, since we do not depend on the defined resources in the specification. In addition, to be able to perform multiple operations on the same resource, we leverage the composition of the proposed behaviours, where to be conformant with a behaviour, the same resource needs to be used.

There exist several methods for generating examples to support the understanding of software~\cite{Buse-Synthesizing-API-usage-examples-2012, Barnaby-Exempla-Gratis-(E.G.)-Code-Examples-for-Free-2020, Gu-CodeKernal-2019,Kim-Adding-Examples-into-Java-Documents-2009, Mar-Recommending-Proper-API-Code-Examples-for-Documentation-Purpose-2011, Montandon-Documenting-APIs-with-examples-Lessons-learned-with-the-APIMiner-platform-2013, Mittal-Generating-examples-for-use-in-tutorial-explanations-using-a-subsumption-based-classifier-1994, Gerdes-Understanding-Formal-Specifications-through-Good-Examples-2018, Holmes-Approximate-Structural-Context-Matching:-An-Approach-to-Recommend-Relevant-Examples-2006, Moreno-How-Can-I-Use-This-Method-2015, Karlsson-Exploring-API-Behaviours-Through-Generated-Examples-2023}. However, most of these approaches rely on white-box information~\cite{Buse-Synthesizing-API-usage-examples-2012, Barnaby-Exempla-Gratis-(E.G.)-Code-Examples-for-Free-2020, Gu-CodeKernal-2019, Kim-Adding-Examples-into-Java-Documents-2009, Mar-Recommending-Proper-API-Code-Examples-for-Documentation-Purpose-2011, Montandon-Documenting-APIs-with-examples-Lessons-learned-with-the-APIMiner-platform-2013, Holmes-Approximate-Structural-Context-Matching:-An-Approach-to-Recommend-Relevant-Examples-2006, Moreno-How-Can-I-Use-This-Method-2015}. Two approaches which do not require white-box information to generate examples are Gerdes et al.~\cite{Gerdes-Understanding-Formal-Specifications-through-Good-Examples-2018} and Karlsson et al~\cite{Karlsson-Exploring-API-Behaviours-Through-Generated-Examples-2023}.
Gerdes et al. proposed a black-box example generating approach which requires a formal specification of the behaviour of the SUT to be able to generate relevant examples~\cite{Gerdes-Understanding-Formal-Specifications-through-Good-Examples-2018}. The limitation of requiring a formal specification was removed with the approach proposed by Karlsson et al.~\cite{Karlsson-Exploring-API-Behaviours-Through-Generated-Examples-2023}. The approach by Karlsson et al. is able to generate relevant examples in a black-box fashion by using generally defined behaviours---as meta-properties. In this paper, we build on this approach by specialising it in the context of REST APIs. In addition, a specialised context makes the approach more relevant to evaluate and compare to other approaches in that same context---REST APIs test generation in our case. 

In summary, our main difference from previous approaches is to drive test generation with a focus on finding behaviours of the SUT. This produces examples of the behaviours the SUT conforms to, usable to both support understanding and as a source of test generation.

\section{Conclusions}\label{sec:conclusions}

Today, \rests{} are a common way for services to provide functionality, both internally, as part of a larger system, and to external clients. This popularity has been reflected in the research community, with many different test generation and fuzzing approaches proposed. However, there are still challenges in this area. 

In this paper, we have mainly focused on bringing test generation for \rests{} closer to the actual business logic. We have done so by proposing a behaviour-based approach of generating examples and an initial set of CRUD-based behavioural-properties. Our evaluation shows that a behaviour-based approach can provide similar coverage to a SotA search-based method, while at the same time producing less state dependent test cases, due to the focus on behaviours. In addition, the approach provide generated examples valuable to practitioners.

We see this work as a starting point that can be further extended with more behaviours, covering a larger domain of \rests{}---more than CRUD-based APIs. Further, the method can be extended to be more consistent in the search results provided.

Searching for and generating test cases can be time-consuming and require many resources. To be respectful of practitioners' time and resources, we should aim for approaches that make the most of the time spent. In addition, time is an essential factor to enable interactive workflows for practitioners. With this work, we have introduced an approach that automates the creation of relevant examples of API behaviours that can serve as (i) a source for engineers to understand the system and (ii) a source of automated test case generation. We have shown that our approach provides these additional benefits while matching---in a given time span---test coverage of SOTA fuzzing approaches. Users do not have to choose to ``only'' get test cases as output. We can also provide them with means of better understanding their systems---an important part of creating high-quality reliable software systems.

\section*{Acknowledgments}
This work is supported by ABB, the industrial postgraduate school Automation Region Research Academy (ARRAY) funded by The Knowledge Foundation.

\bibliographystyle{acm}
\bibliography{main}

\begin{thebibliography}{10}

\bibitem{AWS-cloud}
{Amazon Web Services}.
\newblock \url{https://aws.amazon.com/}.

\bibitem{Curl}
{Curl}.
\newblock \url{https://curl.se/}.

\bibitem{Evomaster-Blackbox-doc}
{EvoMaster Blackbox documentation}.
\newblock
  \url{https://github.com/EMResearch/EvoMaster/blob/master/docs/blackbox.md}.

\bibitem{Evomaster-url}
{EvoMaster Github}.
\newblock \url{https://github.com/EMResearch/EvoMaster}.

\bibitem{Evomaster-1.4}
{EvoMaster Release 1.4}.
\newblock \url{https://github.com/EMResearch/EvoMaster/releases/tag/v1.4.0}.

\bibitem{Google-cloud}
{Google Cloud}.
\newblock \url{https://cloud.google.com/}.

\bibitem{jacoco-url}
{JaCoCo Java Code Coverage Library}.
\newblock \url{https://www.jacoco.org/jacoco/}.

\bibitem{JSON-url}
{JSON}.
\newblock
  \url{https://www.ecma-international.org/publications-and-standards/standards/ecma-404/}.

\bibitem{Azure-cloud}
{Microsoft Azure}.
\newblock \url{https://azure.microsoft.com/en-us/}.

\bibitem{openapi-url}
{OpenAPI Initiative}.
\newblock \url{https://www.openapis.org/}.

\bibitem{quickrest-old-url}
{QuickREST ICST 2020 Replication Package}.
\newblock
  \url{https://github.com/zclj/replication-packages/tree/master/ICST-2020}.

\bibitem{quick-rest-repository}
{QuickREST repository}.
\newblock \url{https://github.com/zclj/QuickREST}.

\bibitem{Martin-Lopez-Testing-of-RESTful-APIs-Promises-and-Challenges-2022}
{\sc {Alberto Martin-Lopez, Sergio Segura, and Antonio Ruiz-Cortés}}.
\newblock {Testing of RESTful APIs: Promises and Challenges}.
\newblock In {\em 30th ACM Joint European Software Engineering Conference and
  Symposium on the Foundations of Software Engineering (ESEC/FSE ’22)\/}
  (2022).

\bibitem{Arcuri-A-Hitchhikers-guide-to-statistical-tests-2014}
{\sc Arcuri, A., and Briand, L.}
\newblock {A Hitchhiker's guide to statistical tests for assessing randomized
  algorithms in software engineering}.
\newblock {\em Software Testing, Verification and Reliability 24}, 3 (2014),
  219--250.

\bibitem{Arcuri-RESTful-API-Automated-Test-Case-Generation-with-EvoMaster-2019}
{\sc {Arcuri, Andrea}}.
\newblock {RESTful API Automated Test Case Generation with EvoMaster}.
\newblock {\em ACM Trans. Softw. Eng. Methodol. 28}, 1 (jan 2019).

\bibitem{Atlidakis-Pythia-Grammar-Based-Fuzzing-of-REST-APIs-2020}
{\sc {Atlidakis, Vaggelis and Geambasu, Roxana and Godefroid, Patrice and
  Polishchuk, Marina and Ray, Baishakhi}}.
\newblock {Pythia: Grammar-Based Fuzzing of REST APIs with Coverage-guided
  Feedback and Learning-based Mutations}, 2020.

\bibitem{Atlidakis-RESTler-Stateful-REST-API-Fuzzing-2019}
{\sc {Atlidakis, Vaggelis and Godefroid, Patrice and Polishchuk, Marina}}.
\newblock {RESTler: Stateful REST API Fuzzing}.
\newblock In {\em {2019 IEEE/ACM 41st International Conference on Software
  Engineering (ICSE)}\/} (2019), pp.~748--758.

\bibitem{Atlidakis-Checking-Security-Properties-of-Cloud-Service-REST-APIs-2020}
{\sc {Atlidakis, Vaggelis and Godefroid, Patrice and Polishchuk, Marina}}.
\newblock {Checking Security Properties of Cloud Service REST APIs}.
\newblock In {\em {2020 IEEE 13th International Conference on Software Testing,
  Validation and Verification (ICST)}\/} (2020), pp.~387--397.

\bibitem{Barnaby-Exempla-Gratis-(E.G.)-Code-Examples-for-Free-2020}
{\sc Barnaby, C., Sen, K., Zhang, T., Glassman, E., and Chandra, S.}
\newblock {Exempla Gratis (E.G.): Code Examples for Free}.
\newblock In {\em {Proceedings of the 28th ACM Joint Meeting on European
  Software Engineering Conference and Symposium on the Foundations of Software
  Engineering}\/} (2020), {ESEC/FSE 2020}, p.~1353–1364.

\bibitem{Buse-Synthesizing-API-usage-examples-2012}
{\sc Buse, R. P.~L., and Weimer, W.}
\newblock {Synthesizing API usage examples}.
\newblock In {\em {2012 34th International Conference on Software Engineering
  (ICSE)}\/} (2012), pp.~782--792.

\bibitem{Claessen-QuickCheck-2000}
{\sc {Claessen, Koen and Hughes, John}}.
\newblock {QuickCheck: A Lightweight Tool for Random Testing of Haskell
  Programs}.
\newblock {\em SIGPLAN Not. 35}, 9 (sep 2000), 268–279.

\bibitem{Corradini-Automated-black-box-testing-of-nominal-and-error-scenarios-in-RESTful-APIs-2022}
{\sc {Corradini, Davide and Zampieri, Amedeo and Pasqua, Michele and
  Viglianisi, Emanuele and Dallago, Michael and Ceccato, Mariano}}.
\newblock {Automated black-box testing of nominal and error scenarios in
  RESTful APIs}.
\newblock {\em {Software Testing, Verification and Reliability} 32}, 5 (2022),
  e1808.

\bibitem{Ed-douibi-Automatic-Generation-of-Test-Cases-for-REST-APIs-A-Specification-Based-Approach-2018}
{\sc {Ed-douibi, Hamza and Cánovas Izquierdo, Javier Luis and Cabot, Jordi}}.
\newblock {Automatic Generation of Test Cases for REST APIs: A
  Specification-Based Approach}.
\newblock In {\em {2018 IEEE 22nd International Enterprise Distributed Object
  Computing Conference (EDOC)}\/} (2018), pp.~181--190.

\bibitem{Fowler-Microservices}
{\sc Fowler, S.~J.}
\newblock {\em Production-Ready Microservices}.
\newblock {O'Reilly}, 2016.

\bibitem{Gerdes-Understanding-Formal-Specifications-through-Good-Examples-2018}
{\sc {Gerdes, Alex and Hughes, John and Smallbone, Nicholas and Hanenberg,
  Stefan and Ivarsson, Sebastian and Wang, Meng}}.
\newblock {Understanding Formal Specifications through Good Examples}.
\newblock In {\em {Proceedings of the 17th ACM SIGPLAN International Workshop
  on Erlang}\/} ({New York, NY, USA}, 2018), Erlang 2018, Association for
  Computing Machinery, p.~13–24.

\bibitem{Godefroid-Intelligent-REST-API-Data-Fuzzing-2020}
{\sc {Godefroid, Patrice and Huang, Bo-Yuan and Polishchuk, Marina}}.
\newblock {Intelligent REST API Data Fuzzing}.
\newblock In {\em Proceedings of the 28th ACM Joint Meeting on European
  Software Engineering Conference and Symposium on the Foundations of Software
  Engineering\/} (New York, NY, USA, 2020), ESEC/FSE 2020, Association for
  Computing Machinery, p.~725–736.

\bibitem{Godefroid-Differential-Regression-Testing-for-REST-APIs-2020}
{\sc {Godefroid, Patrice and Lehmann, Daniel and Polishchuk, Marina}}.
\newblock {Differential Regression Testing for REST APIs}.
\newblock ISSTA 2020, Association for Computing Machinery, p.~312–323.

\bibitem{Golmohammadi-Testing-RESTful-APIs-A-Survey-2023}
{\sc Golmohammadi, A., Zhang, M., and Arcuri, A.}
\newblock Testing restful apis: A survey.
\newblock {\em ACM Trans. Softw. Eng. Methodol.\/} (aug 2023).

\bibitem{Gu-CodeKernal-2019}
{\sc Gu, X., Zhang, H., and Kim, S.}
\newblock {CodeKernel: A Graph Kernel Based Approach to the Selection of API
  Usage Examples}.
\newblock In {\em {2019 34th IEEE/ACM International Conference on Automated
  Software Engineering (ASE)}\/} (2019), pp.~590--601.

\bibitem{Hatfield-Dodds-Deriving-Semantics-Aware-Fuzzers-from-Web-API-Schemas-X-2021}
{\sc {Hatfield-Dodds, Zac and Dygalo, Dmitry}}.
\newblock {Deriving Semantics-Aware Fuzzers from Web API Schemas}, 2021.

\bibitem{Holmes-Approximate-Structural-Context-Matching:-An-Approach-to-Recommend-Relevant-Examples-2006}
{\sc Holmes, R., Walker, R.~J., and Murphy, G.~C.}
\newblock {Approximate Structural Context Matching: An Approach to Recommend
  Relevant Examples}.
\newblock vol.~32, pp.~952--970.

\bibitem{Karlsson-Exploring-API-Behaviours-Through-Generated-Examples-2023}
{\sc {Karlsson, Stefan and Hughes, John and Jongeling, Robbert and Čaušević,
  Adnan and Sundmark, Daniel}}.
\newblock {Exploring API Behaviours Through Generated Examples}.
\newblock In {\em {arXiv 10.48550/arXiv.2308.15210}\/} (2023).

\bibitem{Karlsson-QuickREST-Property-based-Test-Generation-of-OpenAPI-Described-RESTful-APIs-2020}
{\sc {Karlsson, Stefan and Čaušević, Adnan and Sundmark, Daniel}}.
\newblock {QuickREST: Property-based Test Generation of OpenAPI-Described
  RESTful APIs}.
\newblock In {\em {2020 IEEE 13th International Conference on Software Testing,
  Validation and Verification (ICST)}\/} (2020), pp.~131--141.

\bibitem{Kim-Adding-Examples-into-Java-Documents-2009}
{\sc Kim, J., Lee, S., Hwang, S.-w., and Kim, S.}
\newblock {Adding Examples into Java Documents}.
\newblock In {\em {2009 IEEE/ACM International Conference on Automated Software
  Engineering}\/} (2009), pp.~540--544.

\bibitem{Kim-Automated-Test-Generation-for-REST-APIs-No-Time-to-Rest-Yet-2022}
{\sc {Kim, Myeongsoo and Xin, Qi and Sinha, Saurabh and Orso, Alessandro}}.
\newblock {Automated Test Generation for REST APIs: No Time to Rest Yet}.
\newblock In {\em Proceedings of the 31st ACM SIGSOFT International Symposium
  on Software Testing and Analysis\/} (New York, NY, USA, 2022), ISSTA 2022,
  {Association for Computing Machinery}, p.~289–301.

\bibitem{Kontio-Using-the-focus-group-method-in-software-engineering-obtaining-practitioner-and-user-experiences-2004}
{\sc {Kontio, J. and Lehtola, L. and Bragge, J.}}
\newblock {Using the focus group method in software engineering: obtaining
  practitioner and user experiences}.
\newblock In {\em {Proceedings. 2004 International Symposium on Empirical
  Software Engineering, 2004. ISESE '04.}\/} (2004), pp.~271--280.

\bibitem{Laranjeiro-A-Black-Box-Tool-for-Robustness-Testing-of-REST-Services-2021}
{\sc {Laranjeiro, Nuno and Agnelo, João and Bernardino, Jorge}}.
\newblock {A Black Box Tool for Robustness Testing of REST Services}.
\newblock {\em {IEEE Access} 9\/} (2021), 24738--24754.

\bibitem{Mar-Recommending-Proper-API-Code-Examples-for-Documentation-Purpose-2011}
{\sc Mar, L.~W., Wu, Y.-C., and Jiau, H.~C.}
\newblock {Recommending Proper API Code Examples for Documentation Purpose}.
\newblock In {\em {2011 18th Asia-Pacific Software Engineering Conference}\/}
  (2011), pp.~331--338.

\bibitem{Martin-Lopez-RESTest-Automated-Black-Box-Testing-of-RESTful-Web-APIs-2021}
{\sc {Martin-Lopez, Alberto and Segura, Sergio and Ruiz-Cort\'{e}s, Antonio}}.
\newblock {RESTest: Automated Black-Box Testing of RESTful Web APIs}.
\newblock In {\em Proceedings of the 30th ACM SIGSOFT International Symposium
  on Software Testing and Analysis\/} (2021), ISSTA 2021, p.~682–685.

\bibitem{McLellan-Building-more-usable-APIs-1998}
{\sc McLellan, S., Roesler, A., Tempest, J., and Spinuzzi, C.}
\newblock {Building more usable APIs}.
\newblock vol.~15, pp.~78--86.

\bibitem{Mittal-Generating-examples-for-use-in-tutorial-explanations-using-a-subsumption-based-classifier-1994}
{\sc Mittal, V.~O., and Paris, C.}
\newblock {Generating examples for use in tutorial explanations: using a
  subsumption based classifier}.
\newblock In {\em {In Proceedings of the 11th European Conference on Artificial
  Intelligence}\/} (1994).

\bibitem{Montandon-Documenting-APIs-with-examples-Lessons-learned-with-the-APIMiner-platform-2013}
{\sc Montandon, J.~E., Borges, H., Felix, D., and Valente, M.~T.}
\newblock {Documenting APIs with examples: Lessons learned with the APIMiner
  platform}.
\newblock In {\em {2013 20th Working Conference on Reverse Engineering
  (WCRE)}\/} (2013), pp.~401--408.

\bibitem{Moreno-How-Can-I-Use-This-Method-2015}
{\sc Moreno, L., Bavota, G., Di~Penta, M., Oliveto, R., and Marcus, A.}
\newblock {How Can I Use This Method?}
\newblock In {\em 2015 IEEE/ACM 37th IEEE International Conference on Software
  Engineering\/} (2015), vol.~1, pp.~880--890.

\bibitem{Novick-What-Users-Say-They-Want-in-Documentation-2006}
{\sc Novick, D.~G., and Ward, K.}
\newblock {What Users Say They Want in Documentation}.
\newblock In {\em Proceedings of the 24th Annual ACM International Conference
  on Design of Communication\/} (2006), SIGDOC '06, p.~84–91.

\bibitem{Nykaza-What-Programmers-Really-Want:-Results-of-a-Needs-Assessment-for-SDK-Documentation-2002}
{\sc Nykaza, J., Messinger, R., Boehme, F., Norman, C.~L., Mace, M., and
  Gordon, M.}
\newblock {What Programmers Really Want: Results of a Needs Assessment for SDK
  Documentation}.
\newblock In {\em Proceedings of the 20th Annual International Conference on
  Computer Documentation\/} (2002), SIGDOC '02, p.~133–141.

\bibitem{Piccioni-An-Empirical-Study-of-API-Usability-2013}
{\sc Piccioni, M., Furia, C.~A., and Meyer, B.}
\newblock {An Empirical Study of API Usability}.
\newblock In {\em 2013 ACM / IEEE International Symposium on Empirical Software
  Engineering and Measurement\/} (2013), pp.~5--14.

\bibitem{Robillard-A-field-study-of-API-learning-obstacles-2011}
{\sc Robillard, M.~P., and DeLine, R.}
\newblock {A field study of API learning obstacles}.
\newblock vol.~16, pp.~703--732.

\bibitem{Robillard-What-Makes-APIs-Hard-to-Learn?-Answers-from-Developers-2009}
{\sc {Robillard, Martin P.}}
\newblock {What Makes APIs Hard to Learn? Answers from Developers}.
\newblock vol.~26, pp.~27--34.

\bibitem{Breen-A-Practical-Guide-to-Focus-Group-Research-2006}
{\sc {Rosanna L. Breen }}.
\newblock {A Practical Guide to Focus-Group Research}.
\newblock {\em {Journal of Geography in Higher Education} 30}, 3 (2006),
  463--475.

\bibitem{fielding-REST-2000}
{\sc {R.T. Fielding}}.
\newblock {\em {Architectural Styles and the Design of Network-based Software
  Architectures}}.
\newblock PhD thesis, University of California, Irvine, US, 2000.

\bibitem{Segura-Metamorphic-Testing-of-RESTful-Web-APIs-2018}
{\sc {Segura, Sergio and Parejo, Jos\'{e} A. and Troya, Javier and
  Ruiz-Cort\'{e}s, Antonio}}.
\newblock {Metamorphic Testing of RESTful Web APIs}.
\newblock In {\em Proceedings of the 40th International Conference on Software
  Engineering\/} (New York, NY, USA, 2018), ICSE '18, Association for Computing
  Machinery, p.~882.

\bibitem{Serbout-Web-APIs-Structures-and-Data-Models-Analysis-2022}
{\sc Serbout, S., Lauro, F.~D., and Pautasso, C.}
\newblock {Web APIs Structures and Data Models Analysis}.
\newblock In {\em 2022 IEEE 19th International Conference on Software
  Architecture Companion (ICSA-C)\/} (2022), pp.~84--91.

\bibitem{Shull-Investigating-reading-techniques-for-object-oriented-framework-learning-2000}
{\sc Shull, F., Lanubile, F., and Basili, V.}
\newblock {Investigating reading techniques for object-oriented framework
  learning}.
\newblock vol.~26, pp.~1101--1118.

\bibitem{Stallenberg-Improving-Test-Case-Generation-for-REST-APIs-Through-Hierarchical-Clustering-2021}
{\sc {Stallenberg, Dimitri and Olsthoorn, Mitchell and Panichella, Annibale}}.
\newblock {Improving Test Case Generation for REST APIs Through Hierarchical
  Clustering}.
\newblock In {\em 2021 36th IEEE/ACM International Conference on Automated
  Software Engineering (ASE)\/} (2021), pp.~117--128.

\bibitem{Viglianisi-RESTTESTGEN-Automated-Black-Box-Testing-of-RESTful-APIs-2020}
{\sc {Viglianisi, Emanuele and Dallago, Michael and Ceccato, Mariano}}.
\newblock {RESTTESTGEN: Automated Black-Box Testing of RESTful APIs}.
\newblock In {\em {2020 IEEE 13th International Conference on Software Testing,
  Validation and Verification (ICST)}\/} (2020), pp.~142--152.

\bibitem{Wu-Combinatorial-Testing-of-RESTful-APIs-2022}
{\sc {Wu, Huayao and Xu, Lixin and Niu, Xintao and Nie, Changhai}}.
\newblock {Combinatorial Testing of RESTful APIs}.
\newblock In {\em {44th International Conference on Software Engi- neering
  (ICSE ’22)}\/} (2022).

\bibitem{Zhang-Open-Problems-in-Fuzzing-REST-2022}
{\sc {Zhang, Man and Arcuri, Andrea}}.
\newblock {Open Problems in Fuzzing RESTful APIs: A Comparison of Tools}, 2022.

\bibitem{Zhang-Fuzzing-Microservices-In-Industry-2022}
{\sc {Zhang, Man and Arcuri, Andrea and Li, Yonggang and Xue, Kaiming and Wang,
  Zhao and Huo, Jian and Huang, Weiwei}}.
\newblock {Fuzzing Microservices In Industry: Experience of Applying EvoMaster
  at Meituan}, 2022.

\bibitem{Zhang-Resource-and-dependency-based-test-case-generation-for-RESTful-Web-services-2021}
{\sc {Zhang, Man and Marculescu, Bogdan and Arcuri, Andrea}}.
\newblock {Resource and dependency based test case generation for RESTful Web
  services}.
\newblock {\em Empirical Software Engineering\/} (2021).

\end{thebibliography}

\end{document}